\documentclass[preprint,prd,aps,showpacs,nofootinbib]{revtex4}
\usepackage{inputenc}
\usepackage{epsfig,float}
\usepackage{amsfonts}
\usepackage[hypertex]{hyperref}
\usepackage{amsmath}

\newcommand{\im}{{\rm Im}}
\newcommand{\re}{{\rm Re}}
\newcommand{\be}{\begin{eqnarray}}
\newcommand{\ee}{\end{eqnarray}}
\newcommand{\ba}{\begin{array}}
\newcommand{\ea}{\end{array}}

\begin{document}
\title{The dual parametrization for gluon GPDs}

\author{Kirill M. Semenov-Tian-Shansky}
\affiliation{CPhT, \'{E}cole Polytechnique, CNRS,  91128, Palaiseau, France and \\
LPT,   Universit\'{e} d'Orsay, CNRS, 91404 Orsay, France and \\
St. Petersburg State University, St.Petersburg, 198504, Petrodvoretz,   Russia}
\email[]{Kirill.Semenov@cpht.polytechnique.fr}

\preprint{CPHT-RR005.0110}

\begin{abstract}

 We consider the  application of the dual parametrization
for the case of  gluon GPDs in the nucleon. This provides
opportunities for the more flexible modeling unpolarized gluon GPDs in a nucleon which
in particular contain the invaluable information on the fraction of nucleon spin carried by gluons.
We perform the generalization of Abel transform tomography approach for the case of gluons.
We also discuss the skewness effect in the framework of the dual parametrization.
We strongly suggest to employ the fitting strategies based on the dual parametrization to
extract the information on GPDs from the experimental data.

\end{abstract}

\maketitle

\section{Introduction}
%{cyr\_stsh{@}mail.ru}

Generalized parton distributions (GPDs)
\cite{pioneers,pioneerJi}
are considered to
be a promising tool to study the partonic structure of hadrons
(see Refs. \cite{GPV,Diehl,BelRad,Boffi:2007yc} for recent reviews).
Hard exclusive reactions, which can be described in the theoretical framework
of GPDs, provide us opportunity to access the information on GPDs experimentally.

The importance of GPDs was widely realized in connection with the possibility
to study the total angular momentum of partons in the nucleon. This allows to
address the fundamental question how the total nucleon spin is made up from
the contributions due to quarks and gluons.
The famous Ji sum rule
\cite{pioneerJi}
relates the appropriate Mellin moments of quark and gluon GPDs%
\footnote{For the definition of nucleon GPDs we use the set of conventions employed in \cite{Diehl}. See below.}
 $H^{q,g}(x, \xi,t)$
 and
 $E^{q,g}(x, \xi , t)$
to the fractions
$J^{q,g}$
of the total
angular momentum
({\it i.e.} the sum of parton spin and orbital angular momentum)
carried by quarks of the flavor $q$ and gluons respectively:
\be
&&
\int_{-1}^1 dx \, x \left[ H^{q }(x,\xi, \, t=0)+E^{q }(x,\xi, \, t=0) \right]=2J^{q }\,; \nonumber \\
&&
\int_{0}^1 dx \,  \left[H^{g }(x,\xi, \, t=0)+E^{g }(x,\xi, \, t=0) \right]=2J^{g }\,.
\label{Ji}
\ee

Constraining $J^g$ from the experiment would add an important lacking ``piece'' to
the tricky ``nucleon spin puzzle''.
This require the knowledge of gluon GPDs
$H^g$, $E^g$
as functions of $x$ for fixed values of $\xi$ and $t$.
These GPDs can be most preferably studied in the hard exclusive
electroproduction of $J^{PC}=1^{--}$ mesons ($\rho^0, \omega, \phi$).
For these meson electroproduction channels $H^g$ and $E^g$  make
contributions at leading order in
$1/\mathcal{Q}$ ($\mathcal{Q}^2$ refers to the initial photon virtuality)
and in
 $\alpha_s$.
Hard exclusive electroproduction  of
mesons and particulary  the exclusive electroproduction  of
$\rho^0$
is now in focus of intensive experimental investigations
(see {\it e.g.} \cite{Levy:2009gy}).
The important set of  data have been already published
\cite{Morrow:2008ek}.
Future experiments will provide even more precise data over a
broader phase space.
Thus, further theoretical development of GPD formalism that would allow the
interpretation of this new experimental information is highly demanded.

Unfortunately, the problem of GPDs extraction from the data is complicated
by the fact that GPDs depend on several variables
(longitudinal momentum fraction of partons $x$, skewness parameter $\xi$, momentum transfer
squared $t$ and the factorization scale). Moreover, only the integral convolutions of GPDs
with certain convolution kernels rather than GPD themselves enter the observable quantities.
Thus, in order to extract the information on GPDs from the data, one of necessity has to rely upon
different phenomenologicaly motivated parametrizations of GPDs and ingenious
fitting procedures for the observable quantities.
One of the most popular parametrizations of GPDs is
the famous Radyushkin double distribution Ansatz (RDDA)
\cite{RDDA}
employed in numerous phenomenological applications. In particular,
the specific version of RDDA
%the double distribution Ansatz
was adopted for gluon GPDs in the nucleon
\cite{Goloskokov:2005sd,Goloskokov:2007nt}.

There is the growing confidence that the present inability to describe some aspects
of the available experimental data on the hard exclusive processes may be
due to incomplete or inexact way of modelling GPDs (see {\it e.g.} discussions in
\cite{Kumericki:2007sa,Kumericki:2008di,Kumericki:2009uq}).
A possible alternative way to parameterize GPDs consists in employing of the
so-called dual parametrization of GPDs
\cite{Polyakov:2002wz}.
In this paper we apply the dual parametrization approach,
which was originally developed for quark GPDs, to the case of gluon GPDs in the nucleon.
We also perform the generalization of Abel transform tomography method
\cite{Tomography}
for the case of gluons. Finally, we discuss some aspects of the fitting strategy
for the hard exclusive processes observables based on the dual parametrization of GPDs.

\section{Basic definitions}

Following the conventions accepted in \cite{Diehl},
the unpolarized gluon GPDs in nucleon
$H^g$,
$E^g$
are defined as the Fourier transform of the matrix element of the nonlocal gluon operator
between the nucleon states according to:
\be
&& \frac{1}{P \cdot n}
\int \frac{d \lambda}{2 \pi}
e^{i \lambda x P \cdot n}
\langle N(p')|
F^{+ \nu} (-\lambda n/2)
F_{\nu}^+ (\lambda n/2)
| N(p) \rangle \nonumber \\ &&
= \frac{1}{2 P \cdot n}
\bar{U}(P+\frac{\Delta}{2})
\left[
H^g(x,\xi,t) n_\mu \gamma^\mu
\right.
%\nonumber \\ &&
\left.
+ \frac{1}{2 m_N} E^g(x,\xi,t) i \sigma^{\mu \nu} n_\mu \Delta_\nu
\right]
U(P-\frac{\Delta}{2})\,.
\label{Def_gluon_GPDs}
\ee
The polarized gluon GPDs in nucleon
$\tilde{H}^g$,
$\tilde{E}^g$
are defined as
\be
&& \frac{-i}{P \cdot n}
\int \frac{d \lambda}{2 \pi}
e^{i \lambda x P \cdot n}
\langle N(p')|
F^{+ \nu} (-\lambda n/2)
\tilde{F}_{\nu}^+ (\lambda n/2)
| N(p) \rangle \nonumber \\ && = \frac{1}{2 P \cdot n}
\bar{U}(P+\frac{\Delta}{2})
\left[
\tilde{H}^g(x,\xi,t) n_\mu \gamma^\mu \gamma^5
\right.
%\nonumber \\ &&
\left.
+ \frac{1}{2 m_N} \tilde{E}^g(x,\xi,t)   \gamma_5 n_\mu \Delta^\mu
\right]
U(P-\frac{\Delta}{2})\,.
\label{Def_gluon_GPDs_polarized}
\ee

In
(\ref{Def_gluon_GPDs}), (\ref{Def_gluon_GPDs_polarized})
 we employ the standard notations:
$n$ is the light-cone direction ($n^2=0$, $P \cdot n \equiv P^+$),
$P= \frac{1}{2} (p+p')$,
$\Delta= p'-p$,
$t= \Delta^2$,
the skewness variable
$\xi$
refers to $\Delta^+=-2 \xi P^+$;
$\tilde{F}^{\alpha \beta} \equiv \frac{1}{2} \epsilon^{\alpha \beta \gamma \delta } F_{\gamma \delta}$
is the dual gluon field strength. Throughout this paper we adopt the light-cone
gauge $A^+=0$, so that the gauge link does not appear in the operators
in definitions (\ref{Def_gluon_GPDs}), (\ref{Def_gluon_GPDs_polarized}).

The definition
(\ref{Def_gluon_GPDs})
differs from other definitions encountered in the literature (see \cite{Diehl}).
For $0\le x \le 1 $:
\be
&& H^g(x,\xi,t)|_{{\rm here}}= H^g(x,\xi,t)|_{ {[3]}}+H^g(-x,\xi,t)|_{ {[3]}}
%\nonumber \\ &&
=
x \left( H^g(x,\xi,t)|_{ {[17]}}-H^g(-x,\xi,t)|_{ {[17]}} \right)\,.
\nonumber  \\ &&
\tilde{H}^g(x,\xi,t)|_{ {\rm here}}= \tilde{H}^g(x,\xi,t)|_{{ [3]}}- \tilde{H}^g(-x,\xi,t)|_{ {  [3]}}
%\nonumber \\ &&
=
x \left( \tilde{H}^g(x,\xi,t)|_{ {[17]}}+\tilde{H}^g(-x,\xi,t)|_{ {[17]}} \right)\,.
\nonumber
\ee
The same relations hold for $E^g$ and $\tilde{E}^g$ respectively.

As the gluon itself is its own antiparticle
gluon GPDs
$H^g(x,\xi,t)$,
$E^g(x,\xi,t)$
defined in
(\ref{Def_gluon_GPDs})
are even functions of $x$:
\be
&&
H^g(x,\xi,t)=H^g(-x,\xi,t)\,;   \ \  E^g(x,\xi,t)=E^g(-x,\xi,t)\,.
\nonumber \\ &&
\ee
Gluon GPDs
$\tilde{H}^g(x,\xi,t)$,
$\tilde{E}^g(x,\xi,t)$
defined in
(\ref{Def_gluon_GPDs_polarized})
are odd functions of $x$:
\be
&&
\tilde{H}^g(x,\xi,t)=-\tilde{H}^g(-x,\xi,t)\,; \ \   \tilde{E}^g(x,\xi,t)=-\tilde{E}^g(-x,\xi,t)\,.
\nonumber \\ &&
\ee
Let us stress that in what follows we consider the gluon GPDs in nucleon
(\ref{Def_gluon_GPDs}), (\ref{Def_gluon_GPDs_polarized})
on the interval
$0\le x \le 1$.

In the froward limit gluon GPDs
$H^g$ and $\tilde{H}^g$
reduce  to usual forward gluon distributions  in the nucleon
$g(x)$ and $\Delta g(x)$,
while GPD
$E^g$ and $\tilde{E}^g$
are reduced to unknown gluon distributions, which we denote as
$e^g(x)$ and $\Delta e^g(x)$:
\be
&&
H^g(x,0,0)= x g(x); \ \ \ E^g(x,0,0)= x e^g(x) \,; \nonumber \\ &&
\tilde{H}^g(x,0,0)= x \Delta g(x); \ \ \ \tilde{E}^g(x,0,0)= x \Delta e^g(x)\,.
\ee
Note that the forward gluon distributions
$g(x)$, $\Delta g(x)$
and
$e^g(x)$, $\Delta e^g(x)$
are continued to the negative value of their argument according to:
\be
&&
g(x)=-g(-x)\,; \ \ \ e^g(x)=-e^g(-x)\,; \nonumber \\ &&
\Delta g(x)= \Delta g(-x)\,; \ \ \ \Delta e^g(x)= \Delta e^g(-x)\,.
\ee

The Mellin moments in the momentum fraction $x$ are of major importance in
the GPD approach. According to the polynomiality property of GPDs the
$x$ Mellin moments of gluon GPDs defined in
(\ref{Def_gluon_GPDs}),
(\ref{Def_gluon_GPDs_polarized})
are polynomials of $\xi$.
The coefficients of these polynomials are
related to the form factors of the local twist two gluon operators:
\be
&&
\mathcal{O}_g^{\mu \mu_1...\mu_n \nu}=\mathbf{S} F^{\mu \alpha} i \overleftrightarrow{D}^{\mu_1}...i \overleftrightarrow{D}^{\mu_n}
F_{ \alpha  } ^\nu\,;
\nonumber \\ &&
\tilde{\mathcal{O}}_g^{\mu \mu_1...\mu_n \nu}=\mathbf{S} (-i) F^{\mu \alpha} i \overleftrightarrow{D}^{\mu_1}...i \overleftrightarrow{D}^{\mu_n}
\tilde{F}_{ \alpha}^ \nu\,.
\ee
Here, as usual,
$D^\mu$
is the covariant derivative,
$\mathbf{S}$ denotes symmetrization in all uncontracted Lorentz indices and subtraction of the appropriate traces.
More technically, the polynomially property for unpolarized gluon GPDs means that
%\subsection*{Forward limit}
%
%where $g(x)$ stands for the forward gluon distribution in the nucleon.
%
%\subsection*{Polynomiality property}
for odd $N$:
\be
&&
\int_0^1 dx x^{N-1} H^g(x,\xi,t)= \sum_{k=0 \atop  {\rm even}}^{N+\mod({N,2})} \xi^k h^g_{N,k}(t)\,; \nonumber \\ &&
\int_0^1 dx x^{N-1} E^g(x,\xi,t)= \sum_{k=0 \atop {\rm even}}^{N+\mod({N,2})} \xi^k e^g_{N,k}(t) \ \ \ \
\nonumber \\ &&
 {\rm with} \ \ \  e^g_{N,N+1}(t)=-h^g_{N,N+1}(t)\,.
\label{Polynomiality_Gluons}
\ee
For  the case of  polarized gluon GPDs
for even $N$
\be
&&
\int_0^1 dx x^{N-1} \tilde{H}^g(x,\xi,t)= \sum_{k=0 \atop  {\rm even}}^{N } \xi^k \tilde{h}^g_{N,k}(t)\,; \nonumber \\ &&
\int_0^1 dx x^{N-1} \tilde{E}^g(x,\xi,t)= \sum_{k=0 \atop  {\rm even}}^{N} \xi^k \tilde{e}^g_{N,k}(t)\,.
\label{Polynomiality_Gluons_polarized}
\ee

%\subsection*{Combinations suitable for PW expansion in $t$-channel partial waves}

\section{The dual parametrization for gluon GPDs}

Historically the first non-trivial phenomenological para-metrization of GPDs
was the famous Radyushkin double distribution Ansatz (RDDA)
suggested within the double distribution representation of GPDs \cite{RDDA}.
In particular, a version of RDDA was adopted for the case of gluon GPD
$H^g$
in the nucleon (\ref{Def_gluon_GPDs})
\cite{Goloskokov:2005sd,Goloskokov:2007nt}:
\be
&&
  H^g(x,\xi,t=0)=H_{DD}^g(x,\xi)+ 2 \theta(\xi-|x|)
  %\frac{1}{N_f}
%\frac{1}{2}
  \xi D^g \left(\frac{x}{\xi} \right),
  \nonumber \\ &&
  \label{2componentparam_glue}
\ee
where $D^g$ stands for the gluon $D$-term
\cite{GPV}
and
$H_{DD}^g$
is built as a one dimensional section of a two-variable
double distribution:
\be
&&
H_{DD}^g(x,\xi)= \int_{0}^1 d \beta \,
\int_{-1+ \beta}^{1-\beta} d \alpha \,
%\delta(x- \beta - \alpha \xi) \,
\left\{
   \delta(x-\beta- \alpha \xi)
\right.
%\nonumber \\ &&
\left.
   -
    \delta(x+\beta- \alpha \xi)
  \right\}
h^{(b)} (\beta, \alpha) \beta g(\beta).
\nonumber \\ &&
\label{DDpart}
\ee
The profile function
$h^{(b)}(\beta, \alpha)$
is parameterized as
\begin{equation}
h^{(b)}(\beta, \alpha)=\frac{\Gamma(2b+2)}{2^{2b+1} \Gamma^2(b+1)}
\frac{[(1-|\beta|)^2-\alpha^2]^b}{(1-|\beta|)^{2b+1}}\,.
\label{RadProfile}
\end{equation}
The parameter
$b$
characterizes the strength of
$\xi$
dependence of the resulting GPD. The usual choice for the gluon
case is $b=2$.
This is motivated by the interpretation
of the $\alpha$ dependence like a meson distribution amplitude for hard exclusive processes.
The cases $b = 2$ correspond to the asymptotic behavior of a gluon distribution
amplitude $\sim (1-\alpha^2)^2$.

The achieved theoretical understanding of both theoretical and experimental aspects of GPD physics hints
(see {\it e.g.} \cite{Kumericki:2007sa,Kumericki:2008di,Guidal:2007cw})
at the
necessity to introduce new
%of
%the
GPD parametrizations which should be
%are
 more general and flexible than the
basic form of the RDDA employed in present-day mainstream phenomenology.

The possible alternative way to parameterize GPDs is the
so-called dual parametrization
\cite{Polyakov:2002wz}
(see \cite{Tomography,Polyakov:2007rw,Moiseeva:2008qd,ForwardLikeF_KS,DualVSRad} for the recent development and discussion).
Originally, the dual parametrization was formulated for the case of quark GPDs. Now we are going
to generalize this approach for the gluon case.

In the framework of the dual parametrization GPDs are presented as infinite sums of the
$t$-channel%
\footnote{The $t$-channel refers to the $t$-channel of the hard exclusive electroproduction reaction in question.
{\it E.g.} for the case of DVCS this is hadron pair production $\gamma^* \gamma \rightarrow   N \bar{N}$.}
 Regge exchanges
 \cite{Polyakov:1998ze}.
Let us stress that the term ``dual''  is intended to lay emphasis on the natural
association with the old idea of duality in hadron-hadron
low energy scattering.
%\cite{Dolen:1967zz}.
The essence of the
duality hypothesis for binary scattering amplitudes
\cite{Dolen:1967zz,Alfaro_red_book}
can be summarized as the assumption that the infinite sum over
only the cross-channel Regge exchanges may provide the complete description of the
whole scattering amplitude in a certain domain of kinematical variables.

More technically, in
%the framework of
the dual parametrization
the
 $t$-channel matrix element of the particular non-local light ray operator
$\hat{O}$
 between the hadron states which enters the definition of GPD (see {\it e.g.}
 (\ref{Def_gluon_GPDs}), (\ref{Def_gluon_GPDs_polarized}))
 is presented in the following form:
\be
&&
\langle N(p') N(-p)|\,
\hat{O}\,
| 0 \rangle \sim
\sum_{R_J}
\sum_{{\rm polarization} \atop {\rm of \;} R_J }
\frac{1}{t-M_{R_J}^2}
%\nonumber \\ &&
%
\times
\underbrace{\; \langle N(p') N(-p)|R_J\rangle \; }_{R_J N \bar{N} {\rm \; effective \; vertex} }
\underbrace{ \ \ \langle R_J| \,
\hat{O} \,
| 0 \rangle \ \ }_{ {\rm F. T.\; of \; DA \; of \;} R_J}\,.
\nonumber \\ &&
\label{Matrix_element_as_a_sum_of_RJ}
\ee
The sum in
(\ref{Matrix_element_as_a_sum_of_RJ})
stands over all possible $t$-channel meson resonance exchanges with suitable
quantum numbers and of arbitrary high spin
$J$
and mass
$M_{R_J}$.
For the classification of
$R_J N \bar{N}$
vertices see {\it e.g.} ref.~\cite{SemenovTianShansky:2007hv}.
The distribution amplitude (DA) of the spin-$J$ $t$-channel resonance
occurring in
$ \langle R_J| \,\hat{O} \,| 0 \rangle$
matrix element is expanded in the eigenfunctions of the  ERBL (Efremeov, Radyushkin, Brodsky, Lepage)
evolution equation. For the
 gluon case these are the Gegenbauer polynomials
$C_n^{\frac{5}{2}} \left( z\right)$
\cite{Braun:2003rp}.
The on shell spin sum of spin-$J$ $t$-channel resonance resulting from the
sum over polarization in
(\ref{Matrix_element_as_a_sum_of_RJ})
is expanded in the
$t$-channel partial waves.

Thus,  in the framework of the dual parametrization GPDs have the form of
double expansions in the eigenfunctions of the ERBL kernel and in the $t$-channel
partial waves.
These formal series are then analytically continued to the physical region by means of the cunningly
organized summation procedure.
%%%%%%%%%%%%%%%%%%%%%%%%%%%%%%%%%%%%%%%%%%%%%%%%%%%%%%%
It is worthy to mention that
the dual parametrization of GPDs shares many common
features with the
expansion of GPDs in collinear conformal partial waves
\cite{Belitsky:1997pc}
known in different versions
\cite{Shuvaev:1999fm,Kivel:1999wa,Manashov:2005xp,Mueller:2005ed,Kumericki:2007sa,Kumericki:2009uq}.
The expansion of GPDs in collinear conformal partial waves arises
naturally from the solution of the evolution equations to the leading order
accuracy.
In both cases the angular momentum of
${ \rm SO}(3)$
partial waves expansion in the $t$-channel
partial waves as well as the conformal spin appear as labels.

%%%%%%%%%%%%%%%%%%%%%%%%%%%%%%%%%%%%%%%%%%%%%%%%%%%%%%%%

The important point is that one has to take properly into account the complication introduced
by the fact that the nucleon has spin-$\frac{1}{2}$ .
In order to be able to write down the formal series for the gluon GPDs in the nucleon in the framework of the
dual parametrization it is necessary to point out the combinations of gluon GPDs
$H^g$, $\tilde{H}^g$
and
$E^g$, $\tilde{E}^g$
suitable for the partial wave expansion in the $t$-channel partial waves.
This is an easy task since the Lorentz structure of the Fourier transform
of nucleon matrix element of gluon light cone operators which enter the definitions
(\ref{Def_gluon_GPDs}), (\ref{Def_gluon_GPDs_polarized})
is obviously the same as that  of the quark light cone operator familiar
from the definition of unpolarized and polarized quark GPDs respectively.
Thus, in the complete analogy  the case unpolarized quark GPDs
\cite{Diehl,Diehl:2007jb}
the partial wave expansion in the
$t$-channel partial waves can be written for the electric and magnetic combinations of unpolarized gluon GPDs
\be
&&
H^{g\,(E)}(x,\xi,t)= H^g(x,\xi,t)+\tau  E^g(x,\xi,t)\,;  \ \ \ (\tau \equiv  \frac{t}{4 m_N^2})
\nonumber \\ &&
H^{g\,(M)}(x,\xi,t)= H^g(x,\xi,t)+  E^g(x,\xi,t)\,.
\ee
and for the following combinations of polarized gluon GPDs:
\be
&&
\tilde{H}^g (x,\xi,t)\,;  \nonumber \\ &&
\tilde{H}^{g\,(PS)}(x,\xi,t)= \tilde{H}^g(x,\xi,t)+\tau \tilde{E}^g(x,\xi,t)\,.
\ee

One may come to the same conclusion employing the general method for determining
the invariant amplitudes of a binary scattering process suggested in
\cite{Munczec:63,Alfaro:67}
(see also \cite{Alfaro_red_book}).
Using this method it is also straightforward to check that the electric combination
$H^{g\,(E)}$ and pseudoscalar combination
$\tilde{H}^{g\,(PS)}$
are to be expanded in the Legendre polynomials of
$\cos \theta_t$
$P_l(\cos \theta_t)$
while the magnetic combination
$H^{g\,(M)}$ and
$\tilde{H}^{g}$
are to be expanded in the
derivatives of the Legendre polynomials
$P_l'(\cos \theta_t)$.
Note that to the leading order of $1/\mathcal{Q}$ expansion the
$t$-channel scattering angle
\footnote{$\theta_t$ is defined as the scattering angle
in the center of mass frame of the
$t$-channel  of the hard exclusive electroproduction reaction
({\it e.g.} $ \gamma \gamma^* \rightarrow N \bar{N}$  for the DVCS).
}
is expressed through the kinematical variables as
\be
\cos \theta_t= \frac{1}{\xi \sqrt{1- \frac{4 m_N^2}{t}}}+O \left(  \frac{1}{\mathcal{Q}^2}\right)\,.
\ee

For the electric combination of gluon GPDs
$H^{g\,(E)}$
$J^{PC}=0^{++},\;2^{++},\,...$
intermediate meson states
({\it e.g.} $f_0$, $f_2$)
contribute, while for the magnetic combination
$H^{g\,(M)}$
the contributions of
$J^{PC}= 2^{++}, \; 4^{++}, \, ...$
intermediate meson states
are relevant.
For
$\tilde{H}^{g}$
$J^{PC}=1^{++},\;3^{++},\,...$
intermediate meson states contribute
and, finally, for
$\tilde{H}^{g\, (PS)}$
$J^{PC}=0^{-+},\;2^{-+},\,...$
meson states contribute.

Taking into account all these considerations one can write down the following
partial wave expansion for electric and magnetic combinations
of unpolarized gluon GPDs in the nucleon%
\footnote{Note, that contrary to the case of quark GPDs in nucleon defined in
\cite{DualVSRad}
we do not introduce the factor ``$2$'' in the partial wave expansions for gluon GPDs.}:
\be
&&
H^{g\,(E)}(x,\xi,t)=   \sum_{n=1 \atop \rm odd}^\infty \sum_{l=0 \atop \rm even}^{n+1}
B_{n\,l}^{g \, (E)}(t) \,
\theta \left( 1- \frac{x^2}{\xi^2} \right)
\left( 1- \frac{x^2}{\xi^2} \right)^2 %\nonumber \\ && \times
C_{n-1}^{\frac{5}{2}} \left(  \frac{x }{\xi } \right) \xi P_l \left(  \frac{1 }{\xi } \right)\,;
\nonumber \\ &&
\label{Formal_Series_Gluon_E}
\ee
\be
&&
H^{g\,(M)}(x,\xi,t)= \sum_{n=1 \atop \rm odd}^\infty \sum_{l=0 \atop \rm even}^{n+1}
B_{n\,l}^{g \, (M)}(t) \,
\theta \left( 1- \frac{x^2}{\xi^2} \right)
\left( 1- \frac{x^2}{\xi^2} \right)^2  %\nonumber \\ && \times
C_{n-1}^{\frac{5}{2}} \left(  \frac{x }{\xi } \right)  P_l' \left(  \frac{1 }{\xi } \right)\,.
\nonumber \\ &&
\label{Formal_Series_Gluon_M}
\ee

The partial wave expansion for the polarized  gluon GPDs in the nucleon
reads
\be
&&
\tilde{H}^{g }(x,\xi,t)= \sum_{n=2 \atop \rm even}^\infty \sum_{l=1 \atop \rm odd}^{n+1}
\tilde{B}_{n\,l}^{g}(t) \,
\theta \left( 1- \frac{x^2}{\xi^2} \right)
\left( 1- \frac{x^2}{\xi^2} \right)^2
% \nonumber \\ && \times
C_{n-1}^{\frac{5}{2}} \left(  \frac{x }{\xi } \right) \xi P'_l \left(  \frac{1 }{\xi } \right)\,;
\nonumber \\ &&
\label{Formal_Series_Gluon_tilde_H}
\ee
\be
&&
\tilde{H}^{g\, (PS)}(x,\xi,t)= \sum_{n=2 \atop \rm even}^\infty \sum_{l=0 \atop \rm even}^{n}
\tilde{B}_{n\,l}^{g\,(PS)}(t) \,
\theta \left( 1- \frac{x^2}{\xi^2} \right)
\left( 1- \frac{x^2}{\xi^2} \right)^2
% \nonumber \\ && \times
C_{n-1}^{\frac{5}{2}} \left(  \frac{x }{\xi } \right)  P_l \left(  \frac{1 }{\xi } \right)\,.
\nonumber \\ &&
\label{Formal_Series_Gluon_tilde_HPS}
\ee

In fact, the most non-committal way to understand the partial wave expansions
(\ref{Formal_Series_Gluon_E}), (\ref{Formal_Series_Gluon_M}), (\ref{Formal_Series_Gluon_tilde_H}),
(\ref{Formal_Series_Gluon_tilde_HPS})
is to consider them just as formal series which satisfy the fundamental polynomiality
property of gluon GPDs
(\ref{Polynomiality_Gluons}), (\ref{Polynomiality_Gluons_polarized}).
The summation procedure
\cite{Polyakov:2002wz}
allowing to convert these formal series into
rigorously defined expressions is reviewed in the Appendix~\ref{App_A}.

In what follows we are going to consider in details the properties of unpolarized gluon GPDs
in the framework of the dual parametrization. The summary of results for the case of
polarized gluon GPDs in the nucleon is presented in the Appendix~\ref{App_B}.

Let us first consider the electric combination.
For odd $N$ the $(N-1)$th Mellin moment is indeed the polynomial of $\xi$
of order $N+1$
\be
&&
\int_0^1 dx x^{N-1} H^{g\,(E)} (x,\xi,t)= \sum_{k=0 \atop  {\rm even}}^{N+1} \xi^k h^{g\,(E)}_{N,k}(t) \nonumber \\ &&
=\xi^{N} \sum_{n=1 \atop  {\rm odd}}^N
  \sum_{l=0 \atop  {\rm even}}^{n+1}
  B_{nl}^{g\,(E)}(t)
 \xi P_l \left( \frac{1}{\xi} \right)
 %%%%
% \nonumber \\ && \times
 \frac{n\,\left( 1 + n \right) \,\left( 2 + n \right) \,\left( 3 + n \right) \,\Gamma (\frac{5}{2})\,\Gamma (N)}
  {9 \cdot 2^N\,\Gamma (1 + \frac{-n + N}{2})\,\Gamma (\frac{7}{2} + \frac{-2 + n + N}{2})}\,.
 %%%%
\label{Mellin_m_coeff_def_E}
\ee
The set of coefficients
$h^{g\,(E)}_{N,k}(t)$
can be expressed through the generalized form factors
$B_{nl}^{g \, (E)}$
according to
\be
&&
h^{g\,(E)}_{N,k}(t)  = \sum_{n=1 \atop \rm odd}^N
\sum_{l=0 \atop \rm even}^{n+1}
B_{nl}^{g \, (E)}(t) %\nonumber \\ && \times
%%%%%%%%%%%%%%%%%%%%%%%%%%%%%
(-1)^{\frac{k + l - N - 1}{2}}
\frac{ \Gamma (\frac{2 - k + l + N}{2})}{3 \cdot 2^{k+1}  \Gamma (\frac{1 + k + l - N}{2})\,\Gamma (2 - k + N)}
%%%%%%%%%%%%%%%%%%%%%%%%%%%%%
\nonumber \\ &&
\times
\frac{n\,\left( 1 + n \right) \,\left( 2 + n \right) \,\left( 3 + n \right) \,\Gamma (N)}{\Gamma (\frac{2 - n + N}{2})\,\Gamma (\frac{5 + n + N}{2})}\,.
\nonumber \\ &&
%%%%%%%%%%%%%%%%%%%%%%%%%%%%%
\label{coeff_h_E_gluon}
\ee

For the magnetic combination the $(N-1)$th
Mellin moment ($N$- odd) is the the polynomial of $\xi$
of order $N-1$
\be
&&
\int_0^1 dx x^{N-1} H^{g\,(M)} (x,\xi,t)= \sum_{k=0 \atop \text{even}}^{N-1} \xi^k h^{g\,(M)}_{N,k}(t)= \nonumber \\ &&
%=
\xi^{N} \sum_{n=1 \atop \text{odd}}^N
  \sum_{l=0 \atop \text{even}}^{n+1}
  B_{nl}^{g \, (M) }(t)
 \xi P_l' \left( \frac{1}{\xi} \right)
 %%%%
% \nonumber \\ && \times
 \frac{n\,\left( 1 + n \right) \,\left( 2 + n \right) \,\left( 3 + n \right) \,\Gamma (\frac{5}{2})\,\Gamma (N)}
  {9 \cdot 2^N\,\Gamma (1 + \frac{-n + N}{2})\,\Gamma (\frac{7}{2} + \frac{-2 + n + N}{2})}\,.
 %%%%
 %\nonumber \\ &&
\ee
The corresponding set of coefficients
$h^{g\,(M)}_{N,k}(t)$
is expressed through the generalized form factors
$B_{nl}^{g \, (M)}$
as follows
\be
&&
h^{g\,(M)}_{N,k}(t)=    \sum_{n=1 \atop \rm odd}^N
\sum_{l=0 \atop \rm even}^{n+1}
B_{nl}^{g \, (M)}(t)
%\nonumber \\ && \times
%%%%%%%%%%%%%%%%%%%%%%%%%%%%%
(-1)^{\frac{k + l - N + 1}{2}}
 %%%%%%%%%%%%%%%%%%%%%%%%%%%%%
 \frac{\left( -1 + k - N \right) \,\Gamma (\frac{2 - k + l + N}{2})}{ 3 \cdot 2^{k+1} \Gamma (\frac{1 + k + l - N}{2})\,\Gamma (2 - k + N)}
%%%%%%%%%%%%%%%%%%%%%%%%%%%%%%%%
\nonumber \\ &&
\times
\frac{n\,\left( 1 + n \right) \,\left( 2 + n \right) \,\left( 3 + n \right) \,\Gamma (N)}
  {\Gamma (\frac{2 - n + N}{2})\,\Gamma (\frac{5 + n + N}{2})}\,.
  \nonumber \\ &&
  \label{coeff_h_M_gluon}
\ee

Gluon GPDs $H^{g\, (E,M)}(x,\xi,t)$
are normalized according to:
\be
&& \int_{0}^1 dx   H^{g\, (E)}(x, \xi,t)= M_2^g(t)+\frac{4}{5} (1-\tau) d_1^g(t) \, \xi^2\,; \nonumber \\ &&
\int_{0}^1 dx H^{g\,(M)}(x, \xi,t)=2J^g(t)\,.
\label{normalization_HEM_tdep_glue}
\ee
Here
$M_2^g(t)$
stands for the $t$-dependent momentum fraction carried by gluons
in the nucleon;
$J^g(t)$
denotes the $t$-dependent fraction of angular momentum carried by gluons and
$d_1^g(t)$
is the first coefficient of Gegenbauer expansion of the gluon  $D$-term
(\ref{Dterm_gluons}).

\section{The properties of $H^{g\,(E,\,M)}(x,\xi,t)$ in the dual parametrization}

To sum up the formal series for the electric and magnetic combinations
of gluon GPDs
(\ref{Formal_Series_Gluon_E}),
(\ref{Formal_Series_Gluon_M})
we employ the techniques developed in \cite{Polyakov:2002wz} (see also
discussion in \cite{DualVSRad}). Some of the additional  technical details
specific for the gluon case are presented in the
Appendix~\ref{App_A}.

To proceed with the summation of the formal series
(\ref{Formal_Series_Gluon_E}),
(\ref{Formal_Series_Gluon_M})
we introduce two sets of gluon forward-like functions
$G_{2 \nu}^{g \, (E)}$
and
$G_{2 \nu}^{g \, (M)}$,
whose Mellin moments generate electric and magnetic generalized
form-factors
$B_{nl}^{g\, (E,\,M)}(t)$:
\begin{equation}
 B_{n \, n+1-2 \nu}^{g \, (E,\,M)}(t)= \int_0^1 dx x^n G_{2 \nu}^{g \, (E,\,M)}(x,t)\,.
\label{Bnl_EM_glue}
\end{equation}

The explicit expression for the electric combination of
gluon GPDs
$H^{g \,(E)}$
through the corresponding forward like functions
reads:
\be
&& H^{g \,(E)} (x,\xi,t)= %\nonumber \\ &&
\sum_{\nu=0}^\infty
%\left\{
%\frac{}{2} %\right.
\frac{\xi^{2 \nu}}{2}
\left[
H^{g\, (E) \, ( \nu)} (x,\xi,t)+H^{g\, (E)\, ( \nu)} (-x,\xi,t)
\right]\,  \nonumber \\ &&
+
%2
\sum_{\nu=1}^\infty
\theta \left( 1- \frac{x^2}{\xi^2}\right)
\left( 1- \frac{x^2}{\xi^2}\right)^2
\xi C_{2 \nu-2}^{\frac{5}{2}}
\left(
\frac{x}{\xi}
\right)
B_{2 \nu -1 \; 0}^{g\, (E)}(t)\,.
\nonumber \\ &&
\label{HE_dual_through_Gk}
\ee
Note, that the second term in
(\ref{HE_dual_through_Gk})
is the pure $D$-term contribution.
The result for the magnetic combination of gluon
GPDs
$H^{g \,(M)}$
can be obtained in the similar way applying the differential operator
$
\left(1-
x \frac{\partial}{\partial x}-
\xi \frac{\partial}{\partial \xi}
\right)
$:
\be
&&
%\nonumber \\ &&
H^{g \,(M)} (x,\xi,t)=
\sum_{\nu=0}^\infty
\left[
 \left( 1-
x \frac{\partial}{\partial x}-
\xi \frac{\partial}{\partial \xi}
\right)
%\left\{
%\frac{}{2} %
\right.
%\nonumber \\ &&
\left.
\frac{\xi^{2 \nu}}{2}
\left[
H^{g\, (M) \, ( \nu)} (x,\xi,t)+H^{g\, (M)\, ( \nu)} (-x,\xi,t)
\right]\,
%\right.
\right]\,.
 \nonumber \\ &&
 \label{HM_dual_through_Gk}
\ee
The functions
$H^{g \, (E,\,M) \, ( \nu)}(x, \xi, t)$
which appear in
(\ref{HE_dual_through_Gk})
and
(\ref{HM_dual_through_Gk})
defined for
$-\xi \le x \le 1$
are given by the following integral transformations:
\be
 && H^{g \, (E,\,M) \, ( \nu)}(x,\xi,t) \nonumber \\ && =
\theta(x>\xi)
\frac{1}{\pi}
\int_{y_0}^1 dy %\frac{dy}{y}
\left[
\frac{1}{3}\left(
1-y \frac{\partial}{\partial y}+
\frac{1}{2}y^2 \frac{\partial^2}{\partial y^2}
\right)
G_{2 \nu}^{ (E,\,M)}(y,t)
\right]
%\nonumber \\ && \times
\int_{s_1}^{s_2} ds\, \frac{x_s^{2-2 \nu}(1-s^2)}{\sqrt{2 x_s-x_s^2-\xi^2}}
\nonumber
\\&&
%%%%%%%%%%%%%%%%%%%%%%%%%%%%%%%%%%%%%%%%%%%%%%%%%%%%%%%%%%%%%%%%%%%%%%%
+ \theta( |x|<\xi)
\frac{1}{\pi}
\int_{0}^1 dy
\left[
\frac{1}{3}\left(
1-y \frac{\partial}{\partial y}+
\frac{1}{2}y^2 \frac{\partial^2}{\partial y^2}
\right)
G_{2 \nu}^{ (E, \,M)}(y,t)
\right]
%\right]
%\nonumber \\&&  \times
\left\{
\int_{s_1}^{s_3} ds \frac{x_s^{2-2 \nu}(1-s^2)}{\sqrt{2 x_s-x_s^2-\xi^2}}
\right.
 \nonumber \\ &&
 \left.
-
 %%%
 \frac{\pi}{\xi^{2 \nu}}
\left(
1- \frac{x^2}{\xi^2}
\right)^2 \,
 \right.
% \nonumber \\ &&
 \left.
%\times
\sum_{l=-1}^{2 \nu -3}
C_{2 \nu -l-3}^{\frac{5}{2}}
\left(
\frac{x}{\xi}
\right)
\xi P_l
\left(
\frac{1}{\xi}
\right)
\frac{6 y^{2 \nu-l-2}}{(2 \nu-l)(2 \nu -l+1)}
\right\}
\,, %\nonumber \\
\label{Hk_main}
\ee
with
$P_{-n}(\chi) \equiv P_{n-1}(\chi)$.
Here, as usual,
$x_s= 2 \frac{x- \xi s}{(1-s^2)y}$;
$s_i$, ($i=1,...\,4$)
stand for the four roots of the equation
$2 x_s -x_s^2-\xi^2=0$
(see (\ref{SIroots}) for the definitions)
and $y_0$
is defined in (\ref{Y0}).
The integrals in (\ref{Hk_main}) are well convergent
for the set of the forward like functions with the
following small-$y$ behavior
$G_{2 \nu}(y) \sim  \frac{1}{y^{2 \nu +\alpha}}$
with $\alpha <2$.

\subsection*{Point $\xi=x$}
The important limiting case in which the expressions
(\ref{HE_dual_through_Gk}),
(\ref{HM_dual_through_Gk})
are reduced to much simpler forms is the point $x=\xi$.
\be
&&
H^{g\, (E)}(\xi, \xi, t)=
%\nonumber \\ &&
\frac{2}{3 \pi} \xi \int_{\frac{1-\sqrt{1-\xi^2}}{\xi}}^1
\frac{dy}{y} \sum_{\nu=0}^\infty y^{2 \nu}G_{2 \nu}^{  (E)}(y,t)\
\Biggl[
\frac{1}{\sqrt{\frac{2 y}{\xi}-y^2-1}}
\Biggr]\,;
\nonumber \\ &&
H^{g\, (M)}(\xi, \xi, t)= %\nonumber \\ &&
-\xi^2 \frac{\partial}{\partial \xi}
\frac{2}{3 \pi}  \int_{\frac{1-\sqrt{1-\xi^2}}{\xi}}^1
\frac{dy}{y} \sum_{\nu=0}^\infty y^{2 \nu}G_{2 \nu}^{  (M)}(y,t)\
\Biggl[
\frac{1}{\sqrt{\frac{2 y}{\xi}-y^2-1}}
\Biggr]\,.
\nonumber \\ &&
\ee

\subsection*{Point $\xi=1$}

Another notable limiting case is $\xi=1$:
\be
&& H^{g \, (E)}(x, \xi=1, t)
%=\sum_{n=1 \atop \text{odd}}^\infty
%\sum_{l=0 \atop \text{even}}^{n+1}
%B_{n \,l}^{g \, (E)}(t)
%(1-x^2)^2 C_{n-1}^{\frac{5}{2}}(x) P_l(1)
%
=
\frac{1}{2}
(1-x^2)^2
%\nonumber \\ &&
%
%\times
\sum_{\nu=0 }^\infty
\int_0^1 dy y
\left[
\frac{1}{(1-2xy+y^2)^{\frac{5}{2}}}
+
\frac{1}{(1+2xy+y^2)^{\frac{5}{2}}}
\right.
\nonumber \\ &&
\left.
-2\sum_{j=0 \atop \text{even}}^{2 \nu-2} \,
y^j \, C_j^{\frac{5}{2}}(x)
\right] G_{2 \nu}^{ (E)}(y,t)\,;
\nonumber \\ &&
\ee
and
\be
&& H^{g\,(M)}(x, \xi=1, t) =
\frac{1}{2}
(1-x^2)^2
%\nonumber \\ && \times
\sum_{\nu=0}^\infty
\int_0^1 dy
\left[
\frac{1}{(1-2xy+y^2)^{\frac{5}{2}}}+
\frac{1}{(1+2xy+y^2)^{\frac{5}{2}}}
\right.
\nonumber \\&&
\left.
-2\sum_{j=0 \atop  {\rm even}}^{2 \nu-2} \,
y^j \, C_j^{\frac{5}{2}}(x)
\right] y^{3-2 \nu} \frac{\partial}{\partial y} \left( y^{2 \nu} \frac{\partial}{\partial y} \,
G_{2 \nu}^{(M)}(y,t) \right)\,.
\nonumber \\ &&
\ee

\subsection*{Forward limit and $G_0^{ (E,M)}(x,t)$}
Let us introduce the convenient notations for the combinations of the
$t$-dependent parton densities to which GPDs
$H^{g\, (E,M)}$
are reduced in the limit $\xi \rightarrow 0$:
\be
&& H^{g\, (E)}(x, \xi=0,t)= x g(x,t)+
%\frac{t}{4 m_N^2}
\tau
x e^g(x,t) \equiv x g^{(E)}(x,t)\,; \nonumber \\ &&
H^{g\, (M)}(x, \xi=0,t)= x g(x,t)+ x e^g(x,t) \equiv x g^{(M)}(x,t)\,,
\nonumber \\ &&
\ee
where
$g(x,t)$ and $e^g(x,t)$  stands for the $t$-dependent gluon distributions in the nucleon
($x e^g(x,t) \equiv E^g(x, \xi=0,t)$).
Employing the general results
(\ref{coeff_h_E_gluon}),
(\ref{coeff_h_M_gluon})
one can check that for odd $N$:
\be
&&
\int_0^1 dx x^{N-1} H^{g\, (E)}(x, \xi=0,t)=\int_0^1 dx x^{N} g^{(E)}(x, t)
%\nonumber \\ &&
= \frac{(N+2)(N+3)}{3 (2N+3)} B_{N, \, N+1}^{g\, (E)}(t)\,;
\nonumber \\ &&
\int_0^1 dx x^{N-1} H^{g\, (M)}(x, \xi=0,t)=\int_0^1 dx x^{N} g^{(M)}(x, t)
 %\nonumber \\ &&
 = \frac{(N+1)(N+2)(N+3) }{ 3 (2N+3)} B_{N, \, N+1}^{g\, (M)}(t)\,.
\nonumber \\ &&
\label{MellinM_for_G0}
\ee
Inverting the Mellin moments in
(\ref{MellinM_for_G0})
we express the forward like functions
$G_0^{(E,\,M)}$
related to the $t$-dependent gluon distributions
\be
&&
G_0^{  (E)}(x,t)= 9 x^2 \int_x^1 \frac{dy}{y^3} g^{(E)}(y,t)-3 x \int_x^1 \frac{dy}{y^2} g^{(E)}(y,t)\,;
\nonumber\\ &&
G_0^{  (M)}(x,t)= -\frac{9}{2} x^2 \int_x^1 \frac{dy}{y^3} g^{(M)}(y,t)
%\nonumber\\ &&
+3 x \int_x^1 \frac{dy}{y^2} g^{(M)}(y,t)+
\frac{3}{2} \int_x^1 \frac{dy}{y} g^{(M)}(y,t)\,.
\nonumber \\ &&
\label{G0_EM}
\ee
The normalization is
\be
&&
\int_0^1 dx x G_0^{(E)}(x,t)= \frac{5}{4} M_2^g(t)\,;
\nonumber\\ &&
\int_0^1 dx x G_0^{(M)}(x,t)= \frac{5}{8} 2 J^g(t)\,.
\ee
Thus the information on the fraction of nucleon total angular momentum carried by gluons is encoded in the
magnetic forward like function
$G_0^{(M)}$.

\subsection*{Small $\xi$ expansion}
It is helpful to consider the expansion
of electric and magnetic combinations of nucleon gluon GPDs in powers of
$\xi$
around
$\xi=0$
for fixed
$x$ ($x>\xi$).
For the electric combination
$H^{g\, (E)}$
the corresponding expansion to the order
$\xi^2$
it is given by:
\be
&&
H^{g\, (E)}(x,\xi,t)=
\frac{5}{12} x G_0^{(E)}(x,t)-\frac{1}{6} x^2 \frac{\partial}{\partial x}G_0^{(E)}(x,t) +
%\nonumber \\ &&
\frac{1}{8} \int_x^1 dy \left( \frac{x}{y}\right)^{\frac{3}{2}}G_0^{(E)}(y,t)
\nonumber \\ && +\xi^2 \left[
%
%%%% G0 contribution to \xi^2 term
\frac{1}{32}
\left( -\frac{7}{3x} +x\right)
G_0^{(E)}(x,t)+
\frac{1}{24}(1+x^2) \frac{\partial}{\partial x} G_0^{(E)}(x,t)
%\nonumber  \right. \\&& \left.
+
\frac{1}{24}x(1-x^2) \frac{\partial^2}{\partial x^2} G_0^{(E)}(x,t)
\nonumber  \right. \\&& \left.
+ \int_x^1 dy G_0^{(E)}(y,t)
\left(
 \frac{5}{128} \left( \frac{x}{y}\right)^{\frac{1}{2}}+ \frac{3}{128} \left( \frac{x}{y}\right)^{\frac{3}{2}}
 \right.
 \right.
% \nonumber \\ &&
 \left. \left.
+\frac{1}{y^2} \left(  \frac{3}{128} \left( \frac{y}{x}\right)^{\frac{1}{2}}+ \frac{5}{128} \left( \frac{x}{y}\right)^{\frac{1}{2}}  \right)
\right)
\nonumber  \right. \\&& \left.
%%%% G2 contribution  to \xi^2 term
+\frac{7}{48} x G_2^{(E)}(x,t)-
\frac{1}{24}  x^2 \frac{\partial}{\partial x}G_2^{(E)}(x,t)+
\right.
\nonumber \\ &&
\left.
\int_x^1 dy G_2^{(E)}(y,t)
\left(
\frac{35}{256}
\left(\frac{y}{x}\right)^{\frac{1}{2}}+
\frac{5}{128}
\left(\frac{x}{y}\right)^{\frac{1}{2}}+
%$
\frac{3}{256}
\left(\frac{x}{y}\right)^{\frac{3}{2}}
\right)
\right]
%\nonumber
%\\ &&
+O(\xi^4)\,.
\ee
For the magnetic combination
$H^{g\, (E)}$
the expansion to the order
$\xi^2$
reads
\be
&&
H^{g\, (M)}(x,\xi,t)  =
\frac{1}{8} x G_0^{(M)}(x,t)-
\frac{1}{4} x^2 \frac{\partial}{\partial x} G_0^{(M)}(x,t)
%\nonumber    \\ &&
+
\frac{1}{6} x^3 \frac{\partial^2}{\partial x^2} G_0^{(M)}(x,t)
\nonumber    \\ &&
-
\frac{1}{16} \int_x^1 dy \left( \frac{x}{y}\right)^{\frac{3}{2}}G_0^{(M)}(y,t)
%
%\nonumber \\ &&
+\xi^2
\left[
\frac{1}{16 x} G_0^{(M)}(x,t)+
\frac{1}{32} (1-5 x^2)
\frac{\partial}{\partial x}G_0^{(M)}(x,t)
\right.
\nonumber \\ &&
\left.
-
\frac{1}{8} x(1- x^2)
\frac{\partial^2}{\partial x^2}G_0^{(M)}(x,t)
%\right.
%\nonumber \\ &&
%\left.
-
\frac{1}{24} x^2 (1- x^2)
\frac{\partial^3}{\partial x^3}G_0^{(M)}(x,t)
\right.
\nonumber \\ &&
\left.
%%%%%%%%%%%%%%%%%%%%%%%%%%%%%%%%%%%%%%%
- \int_x^1 dy G_0^{(M)}(y,t)
\left(
 \frac{15}{256} \left( \frac{x}{y}\right)^{\frac{1}{2}}+ \frac{15}{256} \left( \frac{x}{y}\right)^{\frac{3}{2}}
 \right.
 \right.
% \nonumber \\ &&
 \left.
 \left.
+\frac{1}{y^2} \left(  \frac{3}{256} \left( \frac{y}{x}\right)^{\frac{1}{2}}+ \frac{15}{256} \left( \frac{x}{y}\right)^{\frac{1}{2}}  \right)
\right)
\nonumber  \right. \\&& \left.
%%%% G2 contribution  to \xi^2 term
-\frac{5}{48} x G_2^{(M)}(x,t)-
\frac{1}{48}  x^2 \frac{\partial}{\partial x}G_2^{(M)}(x,t)
\right.
%\nonumber \\ &&
\left.
+
\frac{1}{24}  x^3 \frac{\partial^2}{\partial x^2}G_2^{(M)}(x,t)
\right. \nonumber \\ && \left.
-\int_x^1 dy G_2^{(M)}(y,t)
\left(
\frac{35}{512}
\left(\frac{y}{x}\right)^{\frac{1}{2}}+
\frac{15}{256}
\left(\frac{x}{y}\right)^{\frac{1}{2}}
\right.
\right.
%\nonumber \\ &&
\left.
\left.
+
%$
\frac{15}{512}
\left(\frac{x}{y}\right)^{\frac{3}{2}}
\right)
\right]
+
 O(\xi^4)\,.
\ee

\subsection*{Modelling $H^{g\, (E)}(x, \xi, t=0$)}
\begin{figure}[H]
 \begin{center}
 \epsfig{figure=  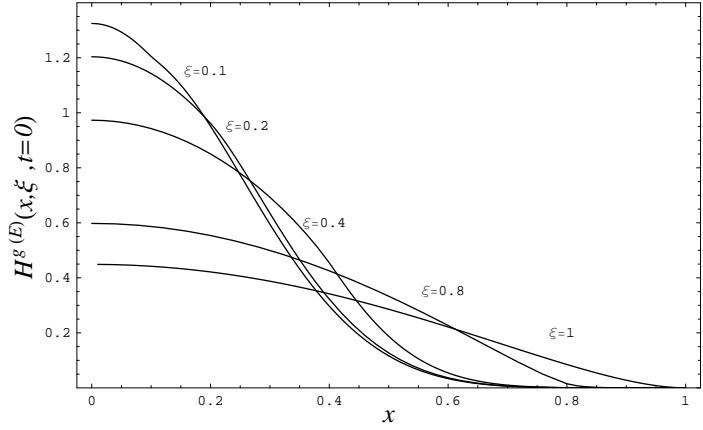 , height=6cm}
  \caption{Electric combinations of nucleon gluon GPDs
$H^{g \,(E)}$,
at
$t=0$
for different values of
$\xi$.
As the numerical input for
forward gluon distributions we use
MRST %NNLO
LO fit
\cite{Martin:2002dr}
for nucleon parton distributions at $Q^2=1 \, \text{GeV}^2$.}
\label{HElg}
\end{center}
\end{figure}

On fig~\ref{HElg}
we show the results of the numerical computation of
$G_0^{(E)}$
contribution into the electric combination of gluon GPDs
$H^{g \,(E)}$
at
$t=0$
for several different values of
$\xi$.
As the numerical input for
forward gluon distributions we used
MRST %NNLO
LO fit
\cite{Martin:2002dr}
for nucleon parton distributions at $Q^2=1 \, \text{GeV}^2$.

\section{Elementary gluon amplitudes and Abel transform tomography}

The typical convolution integral involving gluon GPDs $F^g=\{H^g,\, E^g \}$
relevant for the calculation of hard exclusive $J^{PC}=1^{--}$
meson electroproduction at the leading order in $1/\mathcal{Q}$  and $\alpha_s$
reads \cite{Diehl}:
%See Diehl's review p.175
\be
&&
A^g(\xi,t)= \int_{0}^{1} dx \frac{F^g(x,\xi,t)}{x}
\left[
\frac{1}{\xi-x- i \epsilon}-\frac{1}{\xi+x- i \epsilon}
\right]\,.
\nonumber \\ &&
\ee
We define electric and magnetic elementary amplitudes
\be
&&
A^{g\, (E,M)}(\xi,t)=
%\nonumber \\ &&
\int_{0}^{1} dx \frac{H^{g\, (E,M)} (x,\xi,t)}{x}
\left[
\frac{1}{\xi-x- i \epsilon}-\frac{1}{\xi+x- i \epsilon}
\right]\,.
\nonumber \\ &&
\label{Elementary_ampl_gluon}
\ee
In order to obtain the partial wave expansions of the elementary
amplitudes in the $t$-channel partial wave one has to substitute
the formal series (\ref{Formal_Series_Gluon_E}), (\ref{Formal_Series_Gluon_M})
into the corresponding convolution integrals.
For odd $n \ge 1$
\be
&&
\int_0^\xi dx \left(1- \frac{x^2}{\xi^2}  \right)^2 C_{n-1}^{\frac{5}{2}} \left(  \frac{x }{\xi}  \right)
%\nonumber \\ && \times
  \frac{\xi}{x}
\left[
\frac{1}{\xi-x- i \epsilon}-\frac{1}{\xi+x- i \epsilon}
\right]= \frac{4}{3}\,.
\ee
Thus
\be
&&
A^{g\, (E)}(\xi,t)= \frac{4}{3} \sum_{n=1 \atop \rm odd}^\infty \sum_{l=0 \atop \rm even}^{n+1}
B_{n\,l}^{g\, (E)} P_l  \left(  \frac{1 }{\xi}  \right)\,;
\nonumber \\ &&
A^{g\, (M)}(\xi,t)= \frac{4}{3} \sum_{n=1 \atop \rm odd}^\infty \sum_{l=0 \atop \rm even}^{n+1}
B_{n\,l}^{g\, (M)} \frac{1 }{\xi}  P'_l  \left(  \frac{1 }{\xi}  \right)\,.
\ee
These formal series can be summed exactly as for the case of singlet electric and magnetic
quark GPDs.
The form of the resulting expression actually differs only by a factor $\frac{1}{3}$.
The expressions for gluon electric and magnetic elementary amplitudes
read
\be
&&
A^{g\, (E)}(\xi, t) \nonumber \\ && = \frac{2}{3}
\int_0^1 \frac{dx}{x}
\sum_{\nu=0}^\infty
x^{2 \nu} G_{2 \nu}^{(E)}
%N^{g\, (E)}
(x,t)
%\nonumber \\ && \times
\left[
\frac{1}{\sqrt{1-\frac{2x}{\xi}+x^2}}
+
\frac{1}{\sqrt{1+\frac{2x}{\xi}+x^2}}
-2 \delta_{\nu 0}
\right]\,;
\\ &&
A^{g\, (M)}(\xi, t) \nonumber \\ && = \left( -\xi \frac{\partial}{\partial \xi} \right) \frac{2}{3}
\int_0^1 \frac{dx}{x}
\sum_{\nu=0}^\infty
x^{2 \nu} G_{2 \nu}^{g\, (M)}(x,t)
%\nonumber \\ && \times
\left[
\frac{1}{\sqrt{1-\frac{2x}{\xi}+x^2}}
+
\frac{1}{\sqrt{1+\frac{2x}{\xi}+x^2}}
-2 \delta_{\nu 0}
\right]\,. \nonumber \\ &&
\ee

We introduce electric and magnetic gluon GPD quint-essence functions:
\be
&&
N^{g\, (E)}(x,t)=\sum_{\nu=0}^\infty x^{2\nu}\ G_{2\nu}^{ (E)}(x,t)\,;
%%%%%%%%%%%%%%%%%%%%%%%%%%%%%%%%%%%%%%%%%
\nonumber \\ &&
N^{g \, (M)}(x,t)=\sum_{\nu=0}^\infty x^{2\nu}\ G_{2\nu}^{(M)}(x,t)\,.
\label{QuintessenseEM_gluon}
\ee
The imaginary
parts of gluon electric and magnetic elementary amplitudes then read:
\be
&&
\im A^{g\, (E)}(\xi,t)= \frac{\pi H^{g\, (E)}(\xi, \xi, t)}{\xi} =
%\nonumber \\ &&
\frac{2}{3} \int_{\frac{1-\sqrt{1-\xi^2}}{\xi}}^1
\frac{dx}{x} N^{g\, (E)}(x,t)\
\Biggl[
\frac{1}{\sqrt{\frac{2 x}{\xi}-x^2-1}}
\Biggr]\,;
\nonumber \\ &&
\\ &&
\im  A^{g\, (M)}(\xi,t)= \frac{\pi H^{g\, (M)}(\xi, \xi, t)}{\xi}=
%\nonumber \\ &&
-\frac{2}{3} \int_{\frac{1-\sqrt{1-\xi^2}}{\xi}}^1
dx \, \left\{\frac{\partial }{\partial x}\frac{N^{g \, (M)}(x,t)}{1-\xi x}\right\}\
\Biggl[
\frac{1}{\sqrt{\frac{2 x}{\xi}-x^2-1}}
\Biggr]\,; \nonumber \\ &&
\label{IMMmod}
\ee

The real part of the elementary electric gluon amplitude is given by
\be
&&
\re A^{g \, (E)}(\xi,t)=
\frac{2}{3} \int_0^{\frac{1-\sqrt{1-\xi^2}}{\xi}}
\frac{dx}{x} N^{g\, (E)}(x,t)
\nonumber \\ && \times
\Biggl[
\frac{1}{\sqrt{1-\frac{2 x}{\xi}+x^2}} +
\frac{1}{\sqrt{1+\frac{2 x}{\xi}+x^2}}-\frac{2}{\sqrt{1+x^2}}
\Biggr] \nonumber  \\ &&
  +   \frac{2}{3} \int^1_{\frac{1-\sqrt{1-\xi^2}}{\xi}}
\frac{dx}{x} N^{g \, (E)}(x,t)\
\Biggl[
\frac{1}{\sqrt{1+\frac{2 x}{\xi}+x^2}}-\frac{2}{\sqrt{1+x^2}}
\Biggr] %\nonumber \\ &&
 +
%%%4
2(1-\tau) D^g(t)
\, ,
\ee
where $D^g(t)$ stands for the gluon $D$- form factor
(\ref{GluonDFF}).
Finally, the real part of the magnetic gluon amplitude reads:
\be
&&
\re A^{g \, (M)}(\xi,t)\nonumber \\ && =
- \frac{2}{3} \int_0^{\frac{1-\sqrt{1-\xi^2}}{\xi}}
dx \sqrt{1+x^2}
%\nonumber \\ &&
\left[
\frac{1}{\sqrt{1+x^2-\frac{2x}{\xi}}}+
\frac{1}{\sqrt{1+x^2+\frac{2x}{\xi}}}-
\frac{2}{\sqrt{1+x^2}}
\right]
\nonumber \\&&
\times
\frac{\partial}{\partial x}
\left(
\frac{\sqrt{1+x^2}}{1-x^2} N^{g \, (M)}(x,t)
\right)
\nonumber \\&&
-\frac{2}{3} \int_{\frac{1-\sqrt{1-\xi^2}}{\xi}}^1
dx \sqrt{1+x^2}
\left[
\frac{1}{\sqrt{1+x^2-\frac{2x}{\xi}}}
-\frac{2}{\sqrt{1+x^2}}
\right]
%\nonumber \\&&
\frac{\partial}{\partial x}
\left(
\frac{\sqrt{1+x^2}}{1-x^2} N^{g \, (M)}(x,t)
\right)\,.
\nonumber \\ &&
\ee
The gluon $D$-form factor is defined according to
\be
D^g(t)= \sum_{n=1 \atop  \rm odd}^\infty d^g_n(t)=
%\frac{1}{2}
\int_{-1}^1 dz   \frac{1}{1-z^2}   D^g(z,t)\,,
\label{GluonDFF}
\ee
where $D^g(z,t)$ stand for the gluon $D$-term for which the
following Gegenbauer expansion was adopted in
\cite{GPV}:
\be
&&
D^g(z,t)=
%\nonumber \\ &&
\frac{3}{4}(1-z^2)^2 \left[ d_1^g(t)+d_3^g(t) C_2^{\frac{5}{2}}(z)+ d_5^g(t) C_4^{\frac{5}{2}}(z)+...\right]\,.
\nonumber \\ &&
\label{Dterm_gluons}
\ee
%%%%%%%%%%%%%%%%%%%%%%%%%%%%%%%%%%%%%%%%%%%%%%%%%%%%%%%%%%%%%%%%%%%%%%%%%%%%%%%%%%%%%%%%%%%%%%%%%%%%%%%%%%%%%%%%%%%%%

According to the analysis presented in
\cite{Dispersion,Diehl:2007jb},
the real part of the elementary amplitude
$A^{g\,(E)}$
can be expressed through its imaginary
part with the help a dispersion relation
with one subtraction in the variable
$\omega =\frac{1}{\xi}$
for the fixed value of
$t$:
\be
&&
A^{g \, (E)}(\xi,t)=2 D^g(t)+
%\nonumber \\&&
\frac{1}{\pi} \int_0^1 d \xi'
\left(
\frac{1}{\xi-\xi'-i \epsilon}-
\frac{1}{\xi+\xi'-i \epsilon}
\right)
%\frac{1}{\xi'}
% ?????????????????????????????????????????????????????????????!
\im A^{g \,(E)}(\xi'-i \epsilon,t)\,.
\nonumber \\
\label{Disp_rel}
\ee
The subtraction constant is given by the value of the amplitude at
the non-physical point
$\omega=0\; (\xi = \infty)$.
It is known to be fixed by the
$D$-term and equals
$2 D^g(t)$.

In
\cite{Mueller:2005ed,Kumericki:2007sa,Kumericki:2008di}
it was suggested  to fix the subtraction constant
in the dispersion relation
(\ref{Disp_rel})
in terms of the imaginary part of the corresponding elementary
amplitude assuming special analytical properties
in $j$
of the function
\be
&&
\Phi(j) \equiv \sum_{\nu=0 }^\infty h_{2 \nu+j, \; 2 \nu}^{g\, (E)}(t)
%\nonumber \\&&
=
\int_0^1 dx\, x^{j} \frac{1}{x}
\left[
H^{g\,(E)}(x,x,t)-H^{g\,(E)}(x,0,t)
\right]\,.
\label{SR_dep_on_j}
\ee
For entire odd $j \ge 1$
(\ref{SR_dep_on_j})
defines in the family of sum rules
for the specific combinations of coefficients
$h_{2 \nu+j,\, 2 \nu}^{g\, (E)}$
at powers of
$\xi$
of the Mellin moments of GPD
$H^{g \,(E)}$
(see eq.~(\ref{Mellin_m_coeff_def_E}) for the definition).
For $j=-1$ the integral is divergent since it is usually assumed that
$\frac{1}{x}(H^{g\,(E)}(x,x,t)-H^{g\,(E)}(x,0,t)) \sim 1/x^{\alpha}$ with
$\alpha \sim 1$.
In order to provide the desired expression for the $D$- form factor
(\ref{SR_dep_on_j})
is to be properly analytically continued to $j=-1$.
To make it explicitly, let us assume that
$\frac{1}{x}(H^{g\,(E)}(x,x,t)-H^{g\,(E)}(x,0,t))$
belong to the class of functions with power like behavior for
$x \sim 0$,
which can be presented as the
%following
{\emph{finite}} sums of singular terms
\cite{Gelfand}:
\be
F: \ \ F(x)=
\sum_{r=1}^R
\frac{1}{x^{\alpha_{r}}} f_{r}(x)
\,.
\label{Gelfand_Class}
\ee
We suppose that for all $r=1,\,...R$ $\alpha_r<2$
and
$f_{r}(x)$
are arbitrary functions of
$x$
infinitely differentiable in the vicinity of
$x=0$.
It is also supposed that
$f_{r}(x)$
have zeroes of a sufficiently high order for
$x=1$.
Then the relation for the gluon $D$ form factor  reads as in
\cite{Kumericki:2007sa}:
\be
&&
2 D^g(t)=
\sum_{\nu=0 }^\infty h_{2 \nu-1, \; 2 \nu}^{g\, (E)}(t)
%\nonumber \\&&
=\int_{(0)}^1 dx\,  \frac{1}{x} \cdot  \frac{1}{x}
\left[
H^{g\,(E)}(x,x,t)-H^{g\,(E)}(x,0,t)
\right]\,.
\nonumber \\&&
\label{Reg_atar_g}
\ee
The lower integration limit ``$(0)$'' in
(\ref{Reg_atar_g})
symbolize that we use the so-called analytic (or canonical)
regularization
\cite{Gelfand}:
\be
&&
\int_{(0)}^1 dx \,
 \frac{f(x)}{x^{1+\alpha}}
% \nonumber \\&&
 =
\int_{0}^1 dx \,
 \frac{1}{x^{1+\alpha}}
\left[f(x)-f(0)-x f'(0) \right]
%\nonumber \\&&
 -f(0)\frac{1}{\alpha}  -f'(0)\frac{1}{\alpha-1}\, \ \ \ ( {\rm for} \; \alpha<2)\,.
 \label{Areg} \nonumber \\
\ee

In the framework of the dual parametrization
assuming that
$N^{g\, (E)}(x,t)$
and
 $G_0^{ (E)}(x,t)$
also belong to class
(\ref{Gelfand_Class})
(see discussion in \cite{DualVSRad})
one may rewrite the sum rule
(\ref{Reg_atar_g})
for the gluon $D$-form factor
as
\be
&&
D^g(t)\nonumber \\ && =
\frac{2}{3}
\int_0^1 \frac{dx}{x}G_0^{(E)}(x,t)
\left(
\frac{1}{\sqrt{1+x^2}}
-1
\right)+ %\nonumber \\ &&
\frac{2}{3}
\int_{(0)}^1 \frac{dx}{x}
\left[
N^{g\,(E)}(x,t)-G_0^{(E)}(x,t)
\right]
\frac{1}{\sqrt{1+x^2}}\,.
\nonumber \\ &&
\label{D_form_factor_NQ_reg_gluons}
\ee

Thus the maximal amount of information on gluon GPDs which can be extracted
from the leading order gluon amplitude of hard exclusive electroproduction of
$J^{PC}=1^{--}$
mesons can be quantified in terms of  gluon GPD
quintessence functions
$N^{g\, (E,\,M)}$
and the value of  gluon $D$- form factor $D^g$ or, equivalently,
in terms of $N^{g\, (E,\,M)}$ and $G_0^{(E)}$.
The method of the Abel transform tomography for the dual parametrization of GPDs
suggested in
\cite{Tomography}
allows to invert the integral convolutions for the imaginary parts of the
elementary amplitudes in order to recover GPD quintessence functions.
The generalization of this method for the case of gluon GPDs is straightforward.
The expressions for gluon electric and magnetic GPD quintessence functions
through the imaginary parts of the corresponding elementary amplitudes
(\ref{Elementary_ampl_gluon})
read:
\be
\nonumber
&&  N^{g \, (E)}(x,t)= \nonumber \\&&
-\frac{3}{2 \,\pi} \frac{x(1-x^2)}{(1+x^2)^{\frac{3}{2}}}
\int_{\frac{2x}{1+x^2}}^1 d \xi \,
\frac{1}{(\xi-\frac{2x}{1+x^2})^{\frac{3}{2}}}  %\nonumber \\&&
%\times
\left\{
\frac{1}{\sqrt{\xi}} \, \im A^{g \, (E)}(\xi,t)-
\sqrt{\frac{1+x^2}{2x}} \,
\im A^{g \, (E)}
\left(
\frac{2x}{1+x^2},t
\right)
\right\} \nonumber \\&&
+ \frac{3}{2\, \pi}
\frac{\sqrt{2x}(1+x)}{ \sqrt{1+x^2}} \, \im A^{g \, (E)}
\left(
\frac{2x}{1+x^2},t
\right)\,;  \\&&
N^{g \, (M)}(x,t) %\nonumber \\&&
 =
\frac{3}{2 \pi}
\frac{1-x^2}{\sqrt{1+x^2}}
\int_{\frac{2x}{1+x^2}}^1
\frac{d \xi}{\sqrt{\xi}}
\frac{1}{\sqrt{\xi- \frac{2x}{1+x^2}}} \,
\im
A^{g \, (M)}(\xi,t)\,.
\label{Nmag_abel_gluons}
\ee

%%%%%%%%%%%%%%%%%%%%%%%%%%%%%%%%%%%%%%%%%%%%%%%%%%%%%%%%%%%%%%%%%%%%%%%%%%%%%%%%%%%%%%%%%%%%%%%%%
It is also extremely instructive to consider the small-$\xi$
asymptotic behavior of
$\im A^{g \, (E,M)}(\xi)$
in the framework of the dual parametrization.
Assuming the power law behavior
$
N^{(E,\,M)}(x) \sim \frac{1}{x^\alpha}
$
for gluon electric and magnetic GPD quintessence functions
%$N^{(E,\,M)}(x)$
for small $x$
we obtain for $\xi \sim 0$:
\be
&&
\im A^{g \, (E)}(\xi) \sim \frac{2^{\alpha+1}}{\xi^\alpha} \frac{\Gamma(\frac{1}{2}) \Gamma(\alpha+\frac{1}{2})}{3 \, \Gamma(\alpha+1)}\,;
\nonumber \\ &&
\im A^{g \, (M)}(\xi) \sim \frac{2^{\alpha+1}}{\xi^\alpha} \frac{\Gamma(\frac{1}{2}) \Gamma(\alpha+\frac{1}{2})}{3 \, \Gamma(\alpha)}\,.
\ee

% \ \ \ \  g^{(M)}(x) \sim \frac{1}{x^\alpha}\,.%
The case of particular importance is the so-called minimalist dual model in which
only the contributions of the forward like functions
$G_0^{(E,M)}(x)$
(\ref{G0_EM})
are taken into account.
Let us assume the power law small-$x$ asymptotic behavior for the
electric and magnetic gluon densities
\be
 g^{(E,\,M)}(x) \sim \frac{1}{x^\alpha}\,.%
 \label{power-like_gluons}
\ee
%The phenomenological value of
%$\alpha$
%is $\alpha \approx 1.2$.
For the forward like functions
$G_0^{(E,M)}(x)$
the power law behavior
(\ref{power-like_gluons})
results in
\be
&&
G_0^{(E)}(x) \sim \frac{3(1+2\alpha)}{(\alpha+1)(\alpha+2)} \frac{1}{x^\alpha}\,;
\nonumber \\ &&
G_0^{(M)}(x) \sim \frac{3(1+2\alpha)}{\alpha (\alpha+1)(\alpha+2)} \frac{1}{x^\alpha}\,.
\ee
The contributions of the forward like functions
$G_0^{(E,M)}(x)$
into
$\im A^{g \, (E,M)}$
have the following asymptotic behavior for $\xi \sim 0$:
\be
\im A^{g \, (E,M)}_{G_0}(\xi) \sim
\frac{2^{\alpha+2} \Gamma(\frac{1}{2}) \Gamma(\frac{3}{2}+\alpha)}{  \Gamma(3+\alpha)}
\frac{1}{\xi^{\alpha }}\,.
\label{ImA_Dual_small_xi_asymp_glue}
\ee
%%%%%%%%%%%%%%%%%%%%%%%%%%%%%%%%%%%%%%%%%%%%%
%Thus
%\be
%H^{g \, (E,M)}_{G_0}(\xi, \xi) =\frac{1}{\pi} \xi \, \im A^{g \, (E,M)}_{G_0}(\xi)
%\sim
%\frac{2^{\alpha+2}  \Gamma(\frac{3}{2}+\alpha)}{ \Gamma(\frac{1}{2}) \Gamma(3+\alpha)}
%\frac{1}{\xi^{\alpha-1 }}
%\ee

\section{Skewness effect for small $\xi$ in the dual parametrization}
In this section we discuss some aspects of GPD modelling in the framework of the
dual parametrization.
The important characteristics of the particular GPD model
is the so-called skewness effect (also known as the skewness ratio).
In
\cite{Kumericki:2008di,Kumericki:2009uq}
it was suggested to define the characteristics of the skewness effect for quark and gluon GPDs
$H^{q,\,g}$
for
$t=0$
as ratios of the GPDs on the cross-over line $x=\xi$ to the
appropriate parton distribution functions, to which GPDs are reduced in the
forward limit.
%The argument of PDFs is chosen to be $2 \xi$.
Thus, for quark GPD $H^q$ the skewness ratio is defined according to
\be
&&
r^q= \frac{H^q(x=\xi, \xi, t=0)}{H^q( \xi,  0, t=0)}=\frac{H^q(\xi, \xi ,0)}{q( \xi)}\,,
\nonumber \\ &&
\label{skew_quark}
\ee
while for gluon GPD $H^g$ it reads
\be
&&
r^g= \frac{H^g(x=\xi, \xi , t=0)}{H^g( \xi, 0, t=0)}=\frac{H^g(\xi, \xi ,0)}{ \xi g(\xi)}\,.
\nonumber \\ &&
\label{skew_gluon}
\ee

It is usually assumed that the small-$\xi$ the asymptotic behavior of quark GPD on the
cross over trajectory and the corresponding PDF is governed
by the same leading Regge trajectory:
\be
&&
H^q(\xi,\xi,t=0) = \frac{1}{\pi} \im A^q(\xi,t=0)\;, \ \ \ q(\xi) \ \  \sim \frac{1}{\xi^{a^q(t=0)}}\,.
\nonumber \\ &&
\label{Regge_quarks}
\ee
Similarly, for the gluon case:
\be
&&
\frac{1}{\xi} H^g(\xi,\xi,t=0)= \frac{1}{\pi} \im A^g(\xi,t=0)\;, \ \ \ g(\xi) \ \ \sim \frac{1}{\xi^{a^g(t=0)}}\,.
\nonumber \\ &&
\label{Regge_gluons}
\ee
For $\xi \sim 0$ this makes skewness ratios
(\ref{skew_quark}), (\ref{skew_gluon})
independent of $\xi$.
The physical value of
$a^{q,g}(t=0)$
in the relevant kinematical region is given by:
$a^{q,g}(t=0, \mathcal{Q}^2=4\; {\text GeV}^2) \equiv \alpha^{q,g}=1.1 \div 1.2$.
Obviously, the model without skewness in which
$H^q(x,\xi)=q(x)$
and
$H^g(x,\xi)=xg(x)$
corresponds to
$r^{q,g} = 1$.

%Analogously for the gluon case we suppose

%and the corresponding DIS structure function is governed by the same leading Regge trajectory:

Quark skewness ratio can be directly related to the observable quantities.
Since this point for some time was a rather knotty problem in the literature
we present certain details of the relevant calculations following Ref.~\cite{Kumericki:2009uq}.
It is worth to mention that often in the literature somewhat different quantities are employed to indicate
skewness effect at small $\xi$. They are defined as the following ratios
\cite{Diehl}:
\be
R^q=  \frac{H^q(\xi, \xi ,0)}{q( 2\xi)}\,; \ \ \  R^g=  \frac{H^g(\xi, \xi ,0)}{2 \xi g( 2\xi)}\,.
\label{Def_skewness_Diehl}
\ee
The relation between the two definitions of skewness effect under the assumption
of Regge like asymptotic behavior
(\ref{Regge_quarks}), (\ref{Regge_gluons})
is given by
\be
R^q= 2^{\alpha^q} r^q\,; \ \ \ R^g= 2^{\alpha^g-1} r^g\,.
\ee
The definition
(\ref{Def_skewness_Diehl})
is inspired by the fact that  the observable
ratio of DVCS and DIS cross sections is reduced to $R^q$ rather than $r^q$.
%%%%%%%%%%%%%%%%%%%%%%%%%%%%%%%%%%%%%%%%%%%%%%%%%%%%%%%%%%%%%%%%%%%%%%%%%%%%%%%%%%%%%%%%%%%%%%
%In the LO DVCS analysis the skewness ratio can be expressed
%as the ratio of the singlet unpolarized GPD
%$H_{DVCS}=\frac{4}{9} H^u  + \frac{1}{9}H^d $ on the cross over line to the transverse
%section in H1  kinematics is
%dominated by the contribution stemming from
%and skewness ratio can be expressed as the ratio of this GPD on the cross over trajectory
%
Let us present some details concerning this issue.
In Ref.~\cite{:2007cz} in order to characterize the magnitude of skewness effects
present in the DVCS process for small
$x_{Bj}$ the following ratio
$R$
was introduced as observable quantity:
\be
\mathrm{{R}}(\mathcal{Q}^2, W)= \frac{4 \sqrt{\pi \sigma_{DVCS} \,
%(\mathcal{Q}^2,W)
b(\mathcal{Q}^2)}}{\sigma_T(\gamma^*p \rightarrow
X)\sqrt{1+\rho^2}}\,,
\label{Rdefinition_HERA}
\ee
where $W$
stands for the total invariant energy,
$b(\mathcal{Q}^2)$ is the fitted $t$-slope parameter
($d \sigma_{DVCS}/ dt \sim e^{-b|t|}$).
In the LO analysis the DVCS cross section $\sigma_{DVCS}$  in the small $x_{Bj}$ regime is
governed by quark exchange mechanism
and is known to be
dominated by the contribution due to the imaginary part of the amplitude:
\be
&&
\sigma_{DVCS}(x_{Bj} \sim 0,\mathcal{Q}^2) %\nonumber \\ &&
\simeq
%_{\ \ x_{Bj} \sim 0}
%\frac{\alpha_{e.m.}^2 \pi x_{Bj}^2}{\mathcal{Q}^2} \int_{-\infty}^{t_{min}}
%dt |A_{DVCS}|^2  \nonumber \\ &&=
\frac{\alpha_{e.m.}^2 \pi x_{Bj}^2}{\mathcal{Q}^4}
\left|\im A_{DVCS}(\xi, t=0)\right|^2 \frac{1}{b(\mathcal{Q}^2)}(1+\rho^2)\,.
\nonumber \\ &&
\label{DVCS_cros_sec}
\ee
The imaginary part of the leading order DVCS amplitude is given by
$\im A_{DVCS}(\xi, t=0)= \pi H_{DVCS}(\xi,\xi,t=0)$, where
$H_{DVCS}$
is the combination of the singlet unpolarized quark GPDs
$H_{DVCS}=\frac{4}{9} H^u  + \frac{1}{9}H^d$.
The parameter
$\rho^2$
in
(\ref{Rdefinition_HERA}), (\ref{DVCS_cros_sec})
refers to the small correction due to the real part of the amplitude.
%
%Below, in order to show that to the LO
%$R \simeq H(\xi,\xi,t=0)/H(2\xi,0,t=0)$,
%The expression for the LO DVCS cross section for small
%$x_{Bj}$ reads
Since the virtual photon is assumed to be transversely
polarized, in the case of DVCS it is also to be taken transversely
polarized in the DIS amplitude in
(\ref{Rdefinition_HERA}).
The text book expression for  the transversely polarized
DIS cross section reads:
\be
&&
\sigma_T(\gamma^*p \rightarrow X) \simeq 4 \pi^2 \alpha_{e.m.}
\frac{F_2(x_{Bj},\mathcal{Q}^2)}{\mathcal{Q}^2}
%\nonumber \\ &&
 = 4 \pi^2 \alpha_{e.m.} \frac{x_{Bj} H_{DVCS}(x_{Bj}, \xi=0)}{\mathcal{Q}^2}\,.
%\ \ \ \text{with} \nonumber \\&&
%F_2(x_{Bj},\mathcal{Q}^2)= x_{Bj} \left\{ \frac{4}{9} u_+(x_{Bj})+ \frac{1}{9} d_+(x_{Bj})
%\right\}= x_{Bj} H_{DVCS}(x_{Bj}, \xi=0)
\ee
Now, since for small
$x_{Bj}$
the skewness parameter is given by
$\xi \simeq x_{Bj}/2$,
the definition
(\ref{Rdefinition_HERA})
for the observable
$\mathrm{{R}}$
%sensitive to skewness effect
can be rewritten as
\be
\mathrm{{R}}(\mathcal{Q}^2,W)_{x_{Bj} \sim 0}\simeq %\frac{x_{Bj} \, \im A_{DVCS}(\xi, t=0)}{x_{Bj} H_{DVCS}(x_{Bj},
%\xi=0)}=
\frac{H_{DVCS} \left( \frac{x_{Bj}}{2},\frac{x_{Bj}}{2},t=0 \right)}{H_{DVCS}(x_{Bj},
\xi=0, t=0)}\,.
\ee

\begin{figure*}
 \begin{center}
  \epsfig{figure=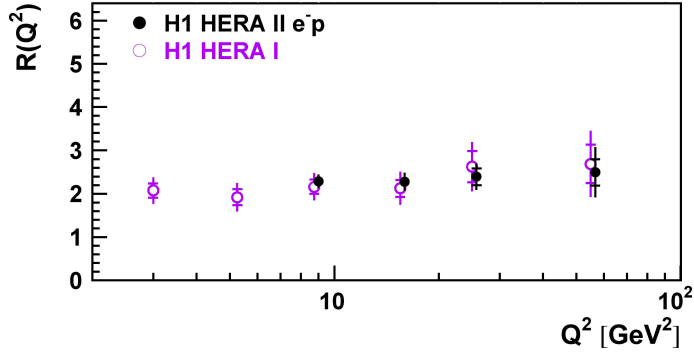, height=5cm}
 \caption{ The observable ratio $\mathrm{R}$
 (\ref{Rdefinition_HERA}),
 shown as a function of $\mathcal{Q}^2$ for fixed
 $W=82$ GeV. This Figure is taken from
 \cite{:2007cz}. }
\label{Fig_HERA}
\end{center}
\end{figure*}

Contrary to the original statement of
Ref.~\cite{:2007cz},
the model without skewness corresponds to
$\mathrm{R}=   2^{\alpha^q} r^q
\simeq 2\,.
$
Thus, as it is emphasized in
\cite{Kumericki:2008di,Kumericki:2009uq},
the results of the experimental measurements of
$\mathrm{R}$
performed by H1 Collaboration
\cite{:2007cz}
presented on Figure~\ref{Fig_HERA}
unambiguously allude to no skewness effect
for small
$x_{Bj}$.

%%%%%%%%%%%%%%%%%%%%%%%%%%%%%%%%%%%%%%%%%%%%%%%%%%%%%%%%%%%%%%%%%%%%%%%%%
Let us now discuss skewness effect in the framework of the dual parametrization.
Employing the results of the previous section and of
\cite{RDDA}
we compute the skewness effect for small
$\xi$
%
%\footnote{Very often in the literature
%(see {\it e.g.} \cite{Kumericki:2009uq})
%the alternative definition of the
%skewness effect is employed: $r^g= \frac{H^g(\xi,\xi)}{H^g(\xi,0)}\approx \frac{1}{2^{\alpha-1}} R^g$}
%(\cite{Diehl})
%for the small-$x_{Bj}$ regime
in the minimalist dual model in which only the contributions of the forward like functions
$Q_0^{(E,M)}(x)$ and $G_0^{(E,M)}(x)$
assuming Regge like behavior
(\ref{Regge_quarks}), (\ref{Regge_gluons})
of input electric and magnetic combinations of parton densities.
The corresponding result reads:
\be
&&
\left. r^{q\,(E,M)}_{Q_0} \equiv {\frac{H^{q \, (E,M)}(\xi, \xi)}{H^{q \, (E,M)}(\xi,0)}} \right|_{\xi \sim 0}
%\nonumber \\ &&
\simeq  \frac{2^{ \alpha^q} \Gamma(\alpha^q+\frac{3}{2})}{ \Gamma( \frac{3}{2}) \Gamma(2+\alpha^q)}
\approx 3/2 \ \ \ {\rm for} \ \ \alpha^q \approx 1\,;
\label{Skewness_effect_Quarks}
\ee
\be
&&
r^{g\,(E,M)}_{G_0} \equiv \left. \frac{H^{g \, (E,M)}(\xi, \xi)}{H^{g \, (E,M)}(\xi,0)} \right|_{\xi \sim 0}
%\nonumber \\ &&
\simeq \frac{2^{ \alpha^g+1} \Gamma(\alpha^g+\frac{3}{2})}{ \Gamma( \frac{3}{2}) \Gamma(3+\alpha^g)}
\approx 1 \ \ \  {\rm for} \ \ \alpha^g \approx 1\,.
\label{Skewness_effect_Gluons}
\ee

It is extremely instructive to compare the skewness effect
(\ref{Skewness_effect_Gluons})
in the minimalist dual model
to that in the commonly used version of RDDA with the
same asymptotic behavior
(\ref{Regge_quarks}), (\ref{Regge_gluons})
of the input parton distributions.
The result for the corresponding skewness ratio  presented {\it e.g.}  in
\cite{Kumericki:2009uq}
reads:
\be
&&
r^{q}_{RDDA} \equiv \frac{H^{q\,}_{DD}(\xi, \xi)}{H^{q }_{DD}(\xi,0)}
\simeq \frac{2^{2 b - \alpha^q} \Gamma(b+\frac{3}{2}) \Gamma(1+b-\alpha^q)}{ \Gamma( \frac{3}{2}) \Gamma(2+2b-\alpha^q)}\,;
\nonumber \\ &&
r^{g}_{RDDA} \equiv \frac{H^{g\,}_{DD}(\xi, \xi)}{H^{g }_{DD}(\xi,0)}
\simeq \frac{2^{2 b+1-\alpha^g} \Gamma(b+\frac{3}{2}) \Gamma(2+b-\alpha^g)}{ \Gamma( \frac{3}{2}) \Gamma(3+2b-\alpha^g)}\,.
\nonumber \\ &&
 \label{Skewness_effect_Gluons_RDDA}
\ee
Comparing this to
(\ref{Skewness_effect_Quarks})
and
(\ref{Skewness_effect_Gluons})
we conclude that both in the quark and gluon cases RDDA with
$b= \alpha^{q,g}$
results in the very same skewness effect for small
$\xi$
as the minimalist dual model with the same input.

Thus, in the quark sector
the minimalist dual model produces the non zero skewness effect for small $\xi$.
It makes this model unpractical for the description
of DVCS at small
$x_{Bj}$
due to systematical
$\sim 50 \%$
overshooting of the available H1 cross section data. At this point
the minimalist dual model has no advantage comparing to the common
version of RDDA plagued by the same problem
(see \cite{Freund:2002qf}
and the discussion in
\cite{Kumericki:2008di,Kumericki:2009uq})
At the same time, in the gluon sector the skewness effect in
the minimalist dual model turns to be very small.
%Unfortunately
%The gluon skewness
%ratio which is relevant for vector meson production experiments
%at small
%$x_{Bj}$
%
It is worth to mention is the growing confidence that the
gluon skewness ratio is also not given by the conformal ratio
(\ref{Skewness_effect_Gluons}) (see discussion in \cite{Kumericki:2009uq}).
The pure LO analysis of hard exclusive $\rho^0$ electroproduction
rather hints that
$r^g<1$.
%can not be accessed in the existing experiments.

In fact one may recognize in
(\ref{Skewness_effect_Gluons})
the well known result obtained in
\cite{Martin:2008tm}
using the Shuvaev integral transform approach
\cite{Shuvaev:1999fm}.
In \cite{Kumericki:2009uq} it was suggested to call it as the conformal ratio
since this is nothing but a Clebsh-Gordan coefficient occurring in the
conformal partial wave expansion.
It is still unclear if the skewness effect
(\ref{Skewness_effect_Gluons})
is an indispensable feature of GPD phenomenology
(see discussion in \cite{Martin:2008tm,Kumericki:2009uq}).

%(the one in which only the contributions of $G_0^{(E,M)}$ are taken into account)
%corresponds to no skewness effect
%(since in the model without skewness $R^g= \frac{\xi g(\xi)}{ 2 \xi g(2 \xi)}= 2^{\alpha-1}$).
%Numerical values of $R^g$ up to $1.6$
%were found in the analysis
%\cite{}.
%However, as pointed in
%\cite{Diehl},
%in the leading $\log 1/x_{Bj}$ approximation
Let us emphasize that
%contrary to the statement made in
%\cite{Kumericki:2009uq}
the skewness ratio in the framework of the dual parametrization
is not with necessity given by
(\ref{Skewness_effect_Gluons}).
In order to alter the skewness ratio  one has to take into account the contribution of
subsequent forward like functions
$Q_{2 \nu}$
($G_{2 \nu}$)
singular enough to
change the small
$\xi$
asymptotic behavior of
$H^{q\,g}(\xi,\xi)$.
For this one has to assume that
$Q_{2 \nu}(x)$, $G_{2 \nu}(x) \sim  1/{x^{2 \nu +\alpha^{q,g}}}$.
%This makes the small $\xi$ asymptotic behavior of GPDs on the cross over trajectory independent of
%
The rigorous way to handle the occurring divergencies of the generalized form
factors
$B_{2 \nu-1 \, 0}$
\be
&&
B_{2 \nu-1 \, 0}^{q \, (E,\,M)}(t)= \int_0^1 dx x^{2 \nu-1} Q_{2 \nu}^{ (E,\,M)}(x,t)\,;
\nonumber \\ &&
B_{2 \nu-1 \, 0}^{g \, (E,\,M)}(t)= \int_0^1 dx x^{2 \nu-1} G_{2 \nu}^{ (E,\,M)}(x,t)\,
\label{Bnl_EM_problematic}
\ee
and of the quark and gluon $D$-form factors was described in details in
\cite{DualVSRad}.

%The small $\xi$ behavior of
%$\im A^{g\, (E)}$
%under the assumptions
%(\ref{power-like_gluons})
%reads
%\be
%&&
%\im
%A_{RDDA}^{g \, (E,M)}(\xi)=
%\frac{\pi}{\xi^2}
%\int_0^{\frac{2 \xi}{1+\xi}} dx \, h^{(b)} \left(x, \frac{\xi-x}{\xi}\right) \, x g^{(E,M)}(x)
%\sim
%\frac{2^{2b+2-\alpha}}{\xi^\alpha}
%\frac{\Gamma(\frac{1}{2}) \Gamma(b+ \frac{3}{2}) \Gamma(2+b-\alpha)}{\Gamma(3+2b-\alpha)}\,.
%\nonumber \\&&
%\label{ImA_DD_small_xi_asymp_glue}
%\ee
%Note that for $\alpha \approx 1$
%the skewness effect $R^{g\,(E,M)}_{DD} \approx 1$ for arbitrary choice of the parameter $b$.
%Thus, changing $b$ does not really alter the skewness effect in RDDA for gluons so that it
%is always very close to the conformal ratio
%(\ref{Skewness_effect_Gluons}).

%%%%%%%%%%%%%%%%%%%%%%%%%%%%%%%%%%%%%%%%%%%%%%%%%%%%%%%%%%%%%%%%%%%%%%%%%%%%%%%%%%%%%

In the remaining part of this section we consider a toy model in order to
briefly sketch the possible fitting strategy for hard exclusive scattering
observables based on the dual parametrization of GPDs. For definiteness we
discuss the case of DVCS at the leading order.

In order to fit the experimental data we have to propose certain Ansatz
for the relevant GPD quintessence functions
$N(x,t)$.
The contribution of the forward like function
$Q_0(x,t)$
into GPD quintessence function is entirely fixed in terms of the $t$-dependent
parton distributions. Since for the moment we discuss DVCS let
$q(x,t)$
be the DVCS combination of $t$-dependent singlet parton distributions
$
q(x,t)=\frac{4}{9} u_+(x,t)+\frac{1}{9} d_+(x,t)\,.
$
%\ee
The observable quantities can be expressed in terms of the
standard  elementary amplitude
\be
&&
A (\xi,t)= \int_{0}^{1} dx H(x,\xi,t)
\left[
\frac{1}{\xi-x- i \epsilon}-\frac{1}{\xi+x- i \epsilon}
\right]\,,
\nonumber \\ &&
\ee
where
$H(x,\xi,t)$
is the DVCS combination of singlet quark GPDs.

The leading singular behavior for
$x \sim 0$
of the
$t$-dependent PDF
$q(x,t)$
is assumed to be determined by the linear Regge trajectory
$a(t) \equiv \alpha+\alpha't$:
\be
q(x,t) \sim c^q \, \frac{1}{x^{a(t)}}\,,
\label{Sing_PDF_q}
\ee
where $c^q>0$ is the numerical constant.
The value of the intercept in the relevant kinematical domain is
$\alpha(\mathcal{Q}^2=4 {\rm GeV}^2)=1.1 \div 1.2$,
The slope parameter
$\alpha'$
can be fixed
{\it e.g.}
with the help of the form factor sum rule
\cite{Guidal:2004nd}:
$\alpha'=1.1 \; \rm{GeV}^{-2}$.

The alluring possibility is to take advantage of the opportunities
provided by the Abel transform tomography method and instead of modelling
GPD quintessence function employ a model for the imaginary part of the
elementary amplitude
$\im A(\xi,t)$.
The real part of the elementary amplitude
$\re A(\xi,t)$
can be then computed with the help of GPD quintessence
restored from $\im A(\xi,t)$ using the tomography method.
The important advantage of this approach is that
the amplitude computed in this way possess proper
analytic properties in $\xi$ and automatically
satisfies the fixed $t$ dispersion relation
with one subtraction in the variable
$\omega =\frac{1}{\xi}$
\cite{Dispersion,Diehl:2007jb}:
\be
&&
A (\xi,t)=4 D (t)+
%\nonumber \\&&
\frac{1}{\pi} \int_0^1 d \xi'
\left(
\frac{1}{\xi-\xi'-i \epsilon}-
\frac{1}{\xi+\xi'-i \epsilon}
\right)
%\frac{1}{\xi'}
% ?????????????????????????????????????????????????????????????!
\im A (\xi'-i \epsilon,t)\,,
\nonumber \\
\label{Disp_rel_quarks}
\ee
since the expression for
$\re A(\xi,t)$
in the dual parametrization
through the GPD quintessence function
is equivalent to the the dispersion relation
(\ref{Disp_rel_quarks}).

Let us introduce the following notations:
$\im A^{{Q_0}}$
and
$\im A^{{N-Q_0}}$
for the contributions to the imaginary part originating from
$Q_0(x,t)$
and
$N(x,t)-Q_0(x,t)$
respectively. Analogously for the
$D$
form factor we introduce the notations
 $D^{Q_0}$ and  $D^{N-Q_0}$.

The leading singular behavior of
$\im A^{{Q_0}}(\xi,t)$
computed from the
$Q_0(x,t)$
corresponding to
$t$-dependent PDF
$q(x,t)$
with the asymptotic behavior
(\ref{Sing_PDF_q})
is
\be
&&
 \im A^{{Q_0}}(\xi,t) \sim c^q \, \frac{2^{a(t)+1}}{\xi^{a(t)}}
  \frac{\Gamma(\frac{1}{2}) \Gamma(a(t)+ \frac{3}{2})}{\Gamma(a(t)+2)} %\nonumber \\ &&
  \equiv C^{Q_0}(t)  \frac{1}{\xi^{a(t)}}\,,
\ee
where we  define the function
\be
&&
C^{Q_0}(t) \equiv c^q \,
\frac{2^{a(t)+1} \Gamma(\frac{1}{2}) \Gamma(a(t)+ \frac{3}{2})}{\Gamma(a(t)+2)}\,
\nonumber \\ &&
 C^{Q_0}(0) \simeq \frac{3 \pi}{2}c^q \ \ \ {\rm for} \ \ \alpha \approx 1\,.
\ee
The contribution of $Q_0$ to the $D$ form factor can be computed
using
\be
D^{Q_0}(t)= \int_0^1   \frac{dx}{x} Q_0(x,t) \left( \frac{1}{\sqrt{1+x^2}}-1\right)\,.
\label{D_Q0}
\ee

The skewness effect in the model which includes only the contribution
of $Q_0$
is given by the conformal ratio
(\ref{Skewness_effect_Quarks}).
In order to make skewness effect satisfy H1 measurements
\cite{:2007cz}
we need to tune the small
$\xi$
asymptotic behavior of
$\im A(\xi,t=0)$.
For this issue let us require the following asymptotic behavior of
$\im A^{N-Q_0}$
for small $\xi$:
\be
&&
\im A^{N-Q_0}(\xi, t=0) \sim     %\frac{\pi}{2} c^q
-\frac{1}{3}  C^{ Q_0}(t=0) \frac{1}{\xi^{a(0)} }
%\nonumber \\ &&
\equiv C^{N-Q_0}(t=0) \frac{1}{\xi^{a(0)} }\,.
\label{How_to_reduce_skewness_eff}
\ee
This choice makes the skewness effect consistent with H1 results.

In order to parameterize the effect of the contribution of
$N(x,t)-Q_0(x,t)$
to the imaginary part of the DVCS amplitude one
may try to employ the following class of functions:
\be
\im A^{{N-Q_0}}(\xi,t)= C^{N-Q_0}(t) \frac{1}{\xi^{a(t)}} (1-\xi )^\beta\,.
\label{Class_0}
\ee
The corresponding truly non-forward contribution to the GPD quintessence function
$N(x,t)-Q_0(x,t)$
can be recovered form
(\ref{Class_0})
employing  the standard Abel
transform tomography procedure and used to compute
the corresponding contribution to the real part
of the elementary amplitude
$\re A^{{N-Q_0}} (\xi,t)$.

%The real part of the amplitude can be computed
%using the GPD quintessence function recovered
%from the imaginary employing the method of .

Finally, the value of the $N-Q_0$ contribution to the
$D$
form factor can be computed rule employing the analyticity
assumptions with the help of inverse momentum sum rule
\cite{Mueller:2005ed,Kumericki:2007sa,Kumericki:2008di}:
\be
&&
D^{{N-Q_0}}(t)= \int_{(0)}^1 \frac{d \xi}{\xi} \im A^{{N-Q_0}}(\xi,t)
%\nonumber \\ &&
=C^{N-Q_0}(t)
{\rm B}(1+\beta,-a(t))\,,
%\frac{\Gamma(1+\beta) \Gamma \left( -a(t)\right) }{  \Gamma \left(1+\beta  -a(t) \right) }
\label{DFF_GAMMA}
\ee
where $\rm B$ is the Euler beta function.

The analytical properties of
$D^{{N-Q_0}}(t)$
in the variable
$t$
require much attention.
The Euler beta function in
(\ref{DFF_GAMMA})
has poles in
$t$
for
$t= \frac{-\alpha+k  }{\alpha'}$
with
$k=0,\,1,\,...\,$.
There is a finite number of
``tachion'' poles at negative values of
$t$
which  do not match with the $t$-channel resonance
exchange picture forming the basis for the
dual parametrization approach.
{\it E.g.} for
$1<\alpha<2$
there are two  ``tachion'' poles at
$t=-\frac{\alpha}{\alpha'}$
and
$t=-\frac{\alpha-1}{\alpha'}$.
This can be seen as an indication that the simple Regge motivated
substitution is inadequate for modelling the $t$-dependence of the DVCS amplitude.
A more sophisticated Ansatz with the non-trivial interplay between the
$\xi$
and
$t$
dependencies is needed for this issue.

The interesting possibility is to employ the
general form
(\ref{Class_0})
assuming the special form of the $t$-dependence of
$C^{N-Q_0}(t)$
in order to get rid of ``tachion''  contributions into  the
$D$ form factor:
\be
&&
C^{N-Q_0}(t)=
%\nonumber \\ &&
C^{N-Q_0}(0) \left(
t+\frac{\alpha}{\alpha'}
\right)
\left(
t+\frac{\alpha-1}{\alpha'}
\right)
\frac{\alpha'^2}{\alpha(\alpha-1)}\,.
\ee
This results in the following expression for the $D$ form factor:
\be
&&
D^{{N-Q_0}}(t)%= \int_{(0)}^1 \frac{d \xi}{\xi} \im A^{{N-Q_0}}(\xi,t)
%\nonumber \\ &&
=C^{N-Q_0}(0)
%B(1+\beta,-a(t))
\frac{1}{\alpha(\alpha-1)}
\frac{\Gamma(1+\beta) \Gamma \left( -a(t)+2 \right) }{ \Gamma \left(1+\beta  -a(t) \right) }\,.
\label{DFF_ispravlennij}
\ee
The value of
$C^{N-Q_0}(0)$
is  necessarily negative in order to reduce the skewness effect for small
$\xi$
(see (\ref{How_to_reduce_skewness_eff})).
It is interesting to note that for
$\beta> a(0) -1$
the value of the
$D^{N-Q_0}$
form factor computed from
(\ref{DFF_ispravlennij})
turns to be negative for
$t=0$%
\footnote{One may check that
the contribution of
$Q_0$
(\ref{D_Q0})
into the
 $D$- form factor is also a small negative number.}.
It is explained in
\cite{Polyakov:1998ze,Kivel:2000fg,GPV}
that the negative value of the
$D$
form factor  is intimately related to the spontaneous breaking of
chiral symmetry in QCD. Some arguments in favor of the negative
sign of the
$D$
form factor are also presented in
\cite{Polyakov:2002yz}
in the framework of the simple model of very large nucleus.
In this case the first coefficient of the Gegenbauer expansion
of the
$D$
form factor can be related to the surface tension of the
hadron medium and its negative sign follows from the requirement
of the mechanical stability of the system.
Thus, at this point our model turns to be consistent
with the achieved theoretical understanding of the underlying physical picture.
Finally it is interesting to note that asymptotic behavior of the
$D$
form factor
$D^{{N-Q_0}}(t)$
for
$-t \rightarrow \infty$
is power like:
\be
D^{{N-Q_0}}(t) \sim (-t)^{-\beta+1}\,.
\ee

This simple exercise shows the power of the analyticity assumptions
allowing to unambiguity fix the subtraction constant in the
dispersion relation
(\ref{Disp_rel_quarks})
in terms of the absorptive part of the amplitude.
The value of the
$D$
form factor
%, whose contribution to the amplitude is supposed to be
%constitutive in the valence region,
turns to be determined by the small
$\xi$
asymptotic behavior of the DVCS cross section.
It is extremely important to  check whether this scenario is consistent with the available
experimental data.

\section{Conclusions}
In this paper we consider the  application of the dual para-metrization approach
to the case of gluon GPDs in a nucleon.
We construct the partial wave expansion for both unpolarized and polarized
gluon GPDs in the nucleon and present the explicit form of the integral transform
allowing to rigorously sum up these formal series.
We present the expressions for the elementary leading order amplitude
entering the description of hard exclusive meson production in the GPD formalism.
We also discuss the generalization of Abel transform tomography approach for the case of gluons.
We argue that the dual parametrization provides
opportunities for the more flexible modeling
%of gluon
GPDs in a nucleon.
% which in particular contain the invaluable information on the fraction of nucleon spin carried by gluons.
We strongly suggest to use the fitting strategies based on the dual parametrization to
extract the information on GPDs from the experimental data.
%%%%%%%%%%%%%%%%%%%%%%%%%%%%%%%%%%%%%%%%%%%%%%%%%%%%%%%%%%%%%%%%%%%%%%%%%%%%%%%%%%%%
Let us also stress that in principle there is no essential difference between
the dual parametrization of GPDs and
the Mellin-Barnes representation  with
the expansion of Gegenbauer moments in the
 in the $t$-channel
 $\mathrm{SO}(3)$
 \cite{Kumericki:2007sa,Kumericki:2009uq}
(see discussion in \cite{Kumericki:2009ji}).
However, the full inversion
formula relating the two parametrizations is still unknown.

\section*{Acknowledgements}
%\appendix
I would especially like to thank Maxim Polyakov for valuable help and
for discussing this article with me.
I am grateful to Dieter M\"{u}ller for numerous illuminating conversations
and for careful reading of the manuscript.
I am also thankful to  Igor Anikin, Alena Moiseeva, Bernard Pire, Lech Szymanowski, Mark Vanderhaeghen and
Samuel Wallon
for many discussions and helpful comments.
I also acknowledge much the partial support from P2I.
%The work was supported by ???.

\setcounter{section}{0}
\setcounter{equation}{0}
\renewcommand{\thesection}{\Alph{section}}
\renewcommand{\theequation}{\thesection\arabic{equation}}

\section{Summing up the formal series for unpolarized gluon GPDs}
\label{App_A}

The method suggested in \cite{Polyakov:2002wz} for the
summation of the formal partial wave expansions of the
type
(\ref{Formal_Series_Gluon_E})
for GPDs in the framework of the dual parametrization consists
in presenting  GPD as the result of  convolution of
a certain convolution kernel with the set of forward like
functions whose Mellin moment generate the generalized form factors
$B_{nl}$.

The explicit form of the corresponding convolution kernel was presented in
\cite{Polyakov:2002wz}
for the case of quark GPD
(see also \cite{DualVSRad} for the refined version of the derivation).
In this appendix we present the summary of relations employed for the derivation
of the integral transformation
(\ref{HE_dual_through_Gk})
expressing GPDs
$H^{g \, (E)}$
through the set of the forward like functions
$G_{2 \nu}^{(E)}$.

%Below we perform the explicit construction of the convolution kernel whose
%convolution with the set of forward like functions
%$Q_{2 \nu}^{g \, (E)}$
%result in the expansion (\ref{Formal_Series_Gluon_E}).

We introduce the common useful variable
\be
z_s=2 \frac{z- \xi s}{(1-s^2)y}\,,
\ee
with
$0<y<1$ and
consider the discontinuity%
\footnote{The discontinuity of the function $f(z)$ is defined as
$
{\rm disc}_{z=x}\,f(z)= \frac{1}{2 \pi i}
\left(
f(x- i0)-f(x+i0)
\right)= \frac{1}{\pi} \im f(x-i0)
$}
\be
&& {\rm disc}_{z=x} \int_{-1}^1 ds \frac{1-s^2}{z_s^{N}} %\nonumber \\ &&
=
\frac{(-1)^{N-1}}{\Gamma(N)}
\int_{-1}^1 ds
(1-s^2)
\delta^{(N-1)}(x_s)  \nonumber \\ && =
(-1)^{N-1}
\theta \left( 1- \frac{x^2}{\xi^2} \right)
\frac{y^N}{2^N \xi^N \Gamma(N)}
%\nonumber \\ && \times
\left.
\left( \frac{\partial}{\partial s} \right)^{N-1}
(1-s^2)^{N+1}
\right|_{s= \frac{x}{\xi}}
\ee
Employing the Rodriguez formula for the Gegenbauer polynomials
$C_n^{\frac{5}{2}}(z)$
\cite{Ryzhik}:
\be
&&
(1-z^2)^2 C_{n-1}^{\frac{5}{2}}(z)=
%\nonumber \\ &&
\frac{(-1)^{n-1}}{2^n \Gamma(n)}
\left(
1+ \frac{5n}{6}+ \frac{n^2}{6}
\right)
\left(
\frac{\partial}{\partial z}
\right)^{n-1}
(1-z^2)^{n+1}
\nonumber \\ &&
\ee
one can derive the following basic relation
\be
&& {\rm disc}_{z=x}
\left(
1+y \frac{\partial}{\partial y}+ \frac{1}{6} y^2 \frac{\partial^2}{\partial y^2}
\right)
%\nonumber \\ && \times
\int_{-1}^1 ds \frac{1-s^2}{z_s^{N}} %\nonumber \\ &&
=
\frac{y^N}{\xi^N}
\theta \left( 1- \frac{x^2}{\xi^2} \right)
\left( 1- \frac{x^2}{\xi^2} \right)^2
 C_{N-1}^{\frac{5}{2}}
 \left(
 \frac{x}{\xi}
 \right)\,.
 \nonumber \\ &&
 \label{basic_disc}
\ee
Now being based on the result (\ref{basic_disc})
we introduce the function
\be
&&
F^{(2 \nu)}(z,y)= %\nonumber \\ &&
\left(
1+y \frac{\partial}{\partial y}+ \frac{1}{6} y^2 \frac{\partial^2}{\partial y^2}
\right)
\int_{-1}^1 ds \xi^{2 \nu} z_s^{2-2\nu}
\frac{1-s^2}{\sqrt{z_s^2-2 z_s+\xi^2}}
\nonumber \\ &&
\ee
whose discontinuity at $z=x$ reads
%$z=x$
\be
&&
{\rm disc}_{z=x} F^{(2 \nu)}(z,y) \nonumber \\ && =
{\rm disc}_{z=x}
\left(
1+y \frac{\partial}{\partial y}+ \frac{1}{6} y^2 \frac{\partial^2}{\partial y^2}
\right)
%\nonumber \\ && \times
\int_{-1}^1 ds
\xi^{2 \nu} \sum_{l=0}^\infty \xi^{l} z_s^{-2 \nu-l+1} (1-s^2) P_l \left( \frac{1}{\xi} \right)
\nonumber \\ &&
=
\theta
\left(
1- \frac{x^2}{\xi^2}
\right)
\left(
1- \frac{x^2}{\xi^2}
\right)^2 {\sum_{l=0}^\infty}^* y^{2 \nu +l-1}  C_{2 \nu +l-2}^{\frac{5}{2}}
 \left(
 \frac{x }{\xi }
\right)
\xi  P_l \left( \frac{1}{\xi} \right)\,.
\nonumber \\ &&
\label{disc_stage0}
\ee
The asterisk in the sum in
(\ref{disc_stage0}) denotes that for
$\nu=0$
the terms with
$l=0$ and $l=1$
are actually absent.

Eq.~(\ref{disc_stage0}) is the natural building block for the desired convolution kernel.
Indeed,
according to the definition of the gluon forward like functions
$G_{2 \nu}^{(E)}(y,t)$ with
$n=2 \nu +l-1$
\be
&&
B_{n \, n+1-2\nu}^{g\,(E)}(t) %\nonumber \\ &&
= \int_0^1 dy y^n
G_{2\nu}^{(E)}(y,t) \ \ \  {\rm with} \ \ \ n \ge 2 \nu-1 \,, \ \ \ {\rm odd.}
\nonumber \\ &&
\label{Bnl_gluons_app}
\ee
Using
(\ref{Bnl_gluons_app})
together with
(\ref{disc_stage0})
it is straightforward to check that the integral convolution
\be
&&
\sum_{\nu =0}^\infty
\int_0^1 dy
\frac{1}{2}
\left\{{\rm disc}_{z=x} F^{(2 \nu)}(z,y)
\right.
%\nonumber \\ &&
\left.
+ {\rm disc}_{z=-x} F^{(2 \nu)}(z,y)
\right\} G_{2 \nu}^{(E)}(y,t)
\ee
results in the formal series
(\ref{Formal_Series_Gluon_E}).

The trick that allows to derive the expression for the convolution kernel
consists in the explicit calculation of the discontinuity of
$F^{(2 \nu)}(z,y)$
at
$z=x$
stemming from the cut at
$1- \sqrt{1-\xi^2}<x_s<1+ \sqrt{1-\xi^2}$
and from possible poles at
$z_s=0$:
\be
&&
\int_0^1 dy \, G_{2 \nu}^{(E)}(y,t) \,{\rm disc}_{z=x} F^{(2 \nu)}(z,y) \nonumber \\&&
= \frac{\xi^{2 \nu}}{\pi}
\int_0^1 dy \, G_{2 \nu}^{(E)}(y,t)
\left(
1+y \frac{\partial}{\partial y}+\frac{1}{6} y^2 \frac{\partial^2}{\partial y^2}
\right)
%\nonumber \\ && \times
\int_{-1}^1 ds \frac{x_s^{2-2 \nu}(1-s^2)}{\sqrt{2x_s-x_s^2-\xi^2}}
\,
\theta(2x_s-x_s^2-\xi^2)
 \nonumber \\&&
-\,
\theta\left(1- \frac{x^2}{\xi^2}\right)
\left(1- \frac{x^2}{\xi^2}\right)^2
%\nonumber \\ && \times
\sum_{l=0}^{2 \nu-3}
C_{2 \nu-l-3}^{\frac{5}{2}}
\left( \frac{x}{\xi} \right)
\xi P_l
\left( \frac{1}{\xi} \right)
\int_0^1 dy y^{2 \nu-l-2} G_{2 \nu}^{(E)}(y,t)\,.
\nonumber \\ &&
 \label{disc_stage1}
\ee

Finally, the expression for the gluon GPD $H^{g\,(E)}$
through the set of the forward like functions
reads
\be
&& H^{g \,(E)}(x,\xi,t)=
\sum_{\nu=0}^\infty
\int_{0}^1 dy \,
\frac{1}{2}
\left\{
{\rm disc}_{z=x} F^{(2 \nu)}(z,y)
\right.
%\nonumber \\ &&
\left.
+
{\rm disc}_{z=-x} F^{(2 \nu)}(z,y)
\right\}
G_{2 \nu}^{(E)}(y,t) \nonumber \\ &&
=\sum_{\nu=0}^\infty
%\left\{
%\frac{}{2} %\right.
\frac{\xi^{2 \nu}}{2}
\left[
H^{g\,(E) \, ( \nu)} (x,\xi,t)+H^{g\,(E)\,( \nu)} (-x,\xi,t)
\right]\,  \nonumber \\ &&
+
%2
\sum_{\nu=1}^\infty
\theta \left( 1- \frac{x^2}{\xi^2}\right)
\left( 1- \frac{x^2}{\xi^2}\right)^2
\xi C_{2 \nu-2}^{\frac{5}{2}}
\left(
\frac{x}{\xi}
\right)
B_{2 \nu -1 \; 0}^{g\, (E)}(t)\,,
%\right.
\nonumber \\ &&
\label{H_dual_through_Gk_App}
\ee
where the functions
$H^{g \,(E)\, ( \nu)}(x, \xi, t)$
defined for
$-\xi \le x \le 1$
are given by the following integral transformations:
\be
 && H^{g \,(E)\, (\nu)}(x,\xi,t)= \nonumber \\ &&
\theta(x>\xi)
\frac{1}{\pi}
\int_{y_0}^1 dy %\frac{dy}{y}
\left[
\frac{1}{3}\left(
1-y \frac{\partial}{\partial y}+
\frac{1}{2}y^2 \frac{\partial^2}{\partial y^2}
\right)
G_{2 \nu}^{(E)}(y,t)
\right]
%\nonumber \\ && \times
\int_{s_1}^{s_2} ds\, \frac{x_s^{2-2 \nu}(1-s^2)}{\sqrt{2 x_s-x_s^2-\xi^2}}
\nonumber
\\&&
%%%%%%%%%%%%%%%%%%%%%%%%%%%%%%%%%%%%%%%%%%%%%%%%%%%%%%%%%%%%%%%%%%%%%%%
+ \theta(
%-\xi<
|x|<\xi)
\frac{1}{\pi}
\int_{0}^1 dy
\left[
\frac{1}{3}\left(
1-y \frac{\partial}{\partial y}+
\frac{1}{2}y^2 \frac{\partial^2}{\partial y^2}
\right)
G_{2 \nu}^{(E)}(y,t)
\right]
%\right]
%\nonumber \\&&  \times
\left\{
\int_{s_1}^{s_3} ds \frac{x_s^{2-2 \nu}(1-s^2)}{\sqrt{2 x_s-x_s^2-\xi^2}}
\right.
\nonumber \\&&
\left.
-
 %%%
 \frac{\pi}{\xi^{2 \nu}}
\left(
1- \frac{x^2}{\xi^2}
\right)^2 \,
\right.
%\nonumber \\ &&
\left. \times
\sum_{l=-1}^{2 \nu -3}
C_{2 \nu -l-3}^{\frac{5}{2}}
\left(
\frac{x}{\xi}
\right)
\xi P_l
\left(
\frac{1}{\xi}
\right)
\frac{6 y^{2 \nu-l-2}}{(2 \nu-l)(2 \nu -l+1)}
\right\}
\,, \nonumber \\
\label{Hk_main_App}
\ee
with
$P_{-n}(\chi) \equiv P_{n-1}(\chi)$.
Note, that in
(\ref{Hk_main_App})
we employ the standard notations
adopted in
\cite{Polyakov:2002wz,DualVSRad}.
Namely,
$x_s= 2 \frac{x- \xi s}{(1-s^2)y}$,
$s_i$, ($i=1,...\,4$)
stand for the four roots of the equation
$2 x_s -x_s^2-\xi^2=0$
given by the following expressions:
\be
\nonumber
&& s_1=\frac{1}{y}
\left(
\mu  - \sqrt{ \left( 1 - x\,y \right) \,\left( 1 + {\mu }^2 \right)-(1 - y^2 )
}
\right); \\&& \nonumber
s_2=\frac{1}{y}
\left(
\mu  + \sqrt{ \left( 1 - x\,y \right) \,\left( 1 + {\mu }^2 \right)-(1 - y^2 )
}
\right);
\\&& \nonumber
s_3=\frac{1}{y}
\left(
\lambda  - \sqrt{ \left( 1 - x\,y \right) \,\left( 1 + {\lambda }^2 \right)-(1 - y^2 )
}
\right);
\\&&
s_4=\frac{1}{y}
\left(
\lambda  + \sqrt{ \left( 1 - x\,y \right) \,\left( 1 + {\lambda }^2 \right)-(1 - y^2 )
}
\right),
 \nonumber \\ &&
\label{SIroots}
\ee
where %we employ the notations:
\be
\mu= \frac{1-\sqrt{1-\xi^2}}{\xi}; \ \ \ \ \lambda=\frac{1}{\mu}\;.
\ee
$y_0$ and
$\frac{1}{y_1}$
are the two solutions of the equation
$s_1=s_2$;
while
$y_1$ and
$\frac{1}{y_0}$
are the two solutions of the equation
$s_3=s_4$;
\be
y_0=
\frac{x\,\left( 1 + {\mu }^2 \right) }{2} +
{\sqrt{  \frac{x^2\,{\left( 1 + {\mu }^2 \right) }^2}{4}-{\mu }^2 }};
\label{Y0}
\ee
\be
y_1=
\frac{x\,\left( 1 + {\lambda }^2 \right) }{2} -
{\sqrt{  \frac{x^2\,{\left( 1 + {\lambda }^2 \right) }^2}{4}-{\lambda }^2 }}.
\ee

\section{Polarized gluon GPDs}
\label{App_B}

As usual, to sum up the formal series for $\tilde{H}^g$
(\ref{Formal_Series_Gluon_tilde_H})
we introduce the set of polarized gluon forward like functions
$\Delta G_{2 \nu} (y,t)$
whose
Mellin moments generate the generalized form factors $\tilde{B}^g_{n\,l}(t)$.
with
$n=2 \nu +l-1$
\be
\tilde{B}_{n \, n+1-2\nu}^{g }(t)= \int_0^1 dy y^n
\Delta G_{2\nu}(y,t) \ \ \  {\rm with} \ \ \ n \ge 2,  \ \ \ {\rm even}.
\nonumber \\ &&
\label{Bnl_gluons_app_polarized}
\ee
The resulting expression for $\tilde{H}^{g }$  through
$\Delta G_{2\nu}(y,t)$
reads
\be
&& \tilde{H}^{g } (x,\xi,t)=
\sum_{\nu=0}^\infty
\left(1-
x \frac{\partial}{\partial x}-
\xi \frac{\partial}{\partial \xi}
\right)
%\left\{
%\frac{}{2} %\right.
\frac{\xi^{2 \nu}}{2}
\left[
\tilde{H} ^{g\, ( \nu)} (x,\xi,t)
\right.
%\nonumber \\ &&
\left.
-\tilde{H}^{g\, ( \nu)} (-x,\xi,t)
\right]\,,  \nonumber \\ &&
\label{Htilde_dual_through_Gk}
\ee
where $\tilde{H} ^{g\, ( \nu)} (x,\xi,t)$
defined for $-\xi<x<1$
is given by
\be
 && \tilde{H}^{g \, ( \nu)}(x,\xi,t) \nonumber \\ && =
\theta(x>\xi)
\frac{1}{\pi}
\int_{y_0}^1 dy %\frac{dy}{y}
\left[
\frac{1}{3}\left(
1-y \frac{\partial}{\partial y}+
\frac{1}{2}y^2 \frac{\partial^2}{\partial y^2}
\right)
\Delta G_{2 \nu}(y,t)
\right]
%\nonumber \\ && \times
\int_{s_1}^{s_2} ds\, \frac{x_s^{2-2 \nu}(1-s^2)}{\sqrt{2 x_s-x_s^2-\xi^2}}
\nonumber
\\&&
%%%%%%%%%%%%%%%%%%%%%%%%%%%%%%%%%%%%%%%%%%%%%%%%%%%%%%%%%%%%%%%%%%%%%%%
+ \theta( |x|<\xi)
\frac{1}{\pi}
\int_{0}^1 dy
\left[
\frac{1}{3}\left(
1-y \frac{\partial}{\partial y}+
\frac{1}{2}y^2 \frac{\partial^2}{\partial y^2}
\right)
\Delta G_{2 \nu} (y,t)
\right]
%\right]
%\nonumber \\&&  \times
\left\{
\int_{s_1}^{s_3} ds \frac{x_s^{2-2 \nu}(1-s^2)}{\sqrt{2 x_s-x_s^2-\xi^2}}
\right.
\nonumber \\ &&
\left.
-
 %%%
 \frac{\pi}{\xi^{2 \nu}}
\left(
1- \frac{x^2}{\xi^2}
\right)^2 \,
\right.
%\nonumber \\ &&
\left. %\times
\sum_{l=-1}^{2 \nu -3}
C_{2 \nu -l-3}^{\frac{5}{2}}
\left(
\frac{x}{\xi}
\right)
\xi P_l
\left(
\frac{1}{\xi}
\right)
\frac{6 y^{2 \nu-l-2}}{(2 \nu-l)(2 \nu -l+1)}
\right\}
\,, \nonumber \\
\label{Hk_main_polarized}
\ee

For even $N$ the polynomiality condition
(\ref{Polynomiality_Gluons_polarized})
require
\be
&&
\int_0^1 dx x^{N-1} \tilde{H}^{g} (x,\xi,t)= \sum_{k=0 \atop \text{even}}^{N} \xi^k \tilde{h}^{g}_{N,k}(t)= \nonumber \\ &&
%=
\xi^{N} \sum_{n=2 \atop  {\rm even}}^N
  \sum_{l=1 \atop  {\rm odd}}^{n+1}
  \tilde{B}_{nl}^{g  }(t)
 \xi P_l' \left( \frac{1}{\xi} \right)
%\nonumber \\ &&
 %%%%
 \frac{n\,\left( 1 + n \right) \,\left( 2 + n \right) \,\left( 3 + n \right) \,\Gamma (\frac{5}{2})\,\Gamma (N)}
  {9 \cdot 2^N\,\Gamma (1 + \frac{-n + N}{2})\,\Gamma (\frac{7}{2} + \frac{-2 + n + N}{2})}\,.
 %%%%
 %\nonumber \\ &&
\ee
The corresponding set of coefficients
$\tilde{h}^{g}_{N,k}(t)$
is expressed through the generalized form factors
$\tilde{B}_{nl}^{g }(t)$
as follows
\be
&&
\tilde{h}^{g }_{N,k}(t)% \nonumber \\ &&
=   \sum_{n=2 \atop \rm even}^N
\sum_{l=1 \atop \rm odd}^{n+1}
\tilde{B}_{nl}^{g }(t) (-1)^{\frac{k + l - N + 1}{2}}
%\nonumber \\ && \times
%%%%%%%%%%%%%%%%%%%%%%%%%%%%%
%
 %%%%%%%%%%%%%%%%%%%%%%%%%%%%%
 \frac{\left( -1 + k - N \right) \,\Gamma (\frac{2 - k + l + N}{2})}{ 3 \cdot 2^{k+1} \Gamma (\frac{1 + k + l - N}{2})\,\Gamma (2 - k + N)}
%%%%%%%%%%%%%%%%%%%%%%%%%%%%%%%%
\nonumber \\ && \times
\frac{n\,\left( 1 + n \right) \,\left( 2 + n \right) \,\left( 3 + n \right) \,\Gamma (N)}
  {\Gamma (\frac{2 - n + N}{2})\,\Gamma (\frac{5 + n + N}{2})}\,.
  \nonumber \\ &&
  \label{coeff_h_gluon_polarized}
\ee

The expression for the polarized forward like function
$\Delta G_0$
through the $t$-dependent polarized gluon density
$\Delta g (y,t)$
reads
\be
&&
\Delta G_0(x,t)= -\frac{9}{2} x^2 \int_x^1 \frac{dy}{y^3} \Delta g (y,t)+3 x \int_x^1 \frac{dy}{y^2} \Delta g (y,t)
%\nonumber \\ &&
+
\frac{3}{2} \int_x^1 \frac{dy}{y} \Delta g (y,t)\,.
\ee

Finally, to sum up the formal series for $\tilde{H}^{g\, (PS)}$
(\ref{Formal_Series_Gluon_tilde_HPS})
the set of polarized gluon forward like functions
$\Delta G_{2 \nu}^{(PS)} (y,t)$
whose
Mellin moments generate the generalized form factors $\tilde{B}^g_{n\,l}(t)$.
with
$n=2 \nu +l $
\be
&&
\tilde{B}_{n \, n-2\nu}^{g\, (PS) }(t)= \int_0^1 dy y^n
\Delta G_{2\nu}^{PS}(y,t) \ \ \  {\rm with} \ \ \ n \ge 2,  \ \ \  {\rm even}.
\nonumber \\ &&
\label{Bnl_gluons_app_polarized_PS}
\ee
The resulting expression for $\tilde{H}^{g }$  through
$\Delta G_{2\nu}$
reads
\be
&& \tilde{H}^{g \, (PS) } (x,\xi,t)=
%\nonumber \\ &&
\sum_{\nu=0}^\infty
%\left\{
%\frac{}{2} %\right.
\frac{\xi^{2 \nu}}{2}
\left[
\tilde{H} ^{g\, (PS)\,  ( \nu)} (x,\xi,t)-\tilde{H}^{g\,(PS) \, ( \nu)} (-x,\xi,t)
\right]\,,  \nonumber \\ &&
\label{Htilde_dual_through_Gk_PS}
\ee
where
$\tilde{H} ^{g\, (PS) \, ( \nu)} (x,\xi,t)$
defined for $-\xi<x<1$
is given by
\be
 && \tilde{H}^{g \,(PS)\,  ( \nu)}(x,\xi,t) \nonumber \\ && =
\theta(x>\xi)
\frac{1}{\pi}
\int_{y_0}^1 dy %\frac{dy}{y}
\left[
\frac{1}{3}\left(
1-y \frac{\partial}{\partial y}+
\frac{1}{2}y^2 \frac{\partial^2}{\partial y^2}
\right)
\Delta G_{2 \nu}^{(PS)}(y,t)
\right] %\nonumber \\ &&  \times
\int_{s_1}^{s_2} ds\, \frac{x_s^{1-2 \nu}(1-s^2)}{\sqrt{2 x_s-x_s^2-\xi^2}}
\nonumber
\\&&
%%%%%%%%%%%%%%%%%%%%%%%%%%%%%%%%%%%%%%%%%%%%%%%%%%%%%%%%%%%%%%%%%%%%%%%
+ \theta(  |x|<\xi)
\frac{1}{\pi}
\int_{0}^1 dy
\left[
\frac{1}{3}\left(
1-y \frac{\partial}{\partial y}+
\frac{1}{2}y^2 \frac{\partial^2}{\partial y^2}
\right)
\Delta G_{2 \nu}^{(PS)} (y,t)
\right]
%\right]
%\nonumber \\&&  \times
\left\{
\int_{s_1}^{s_3} ds \frac{x_s^{1-2 \nu}(1-s^2)}{\sqrt{2 x_s-x_s^2-\xi^2}}
\right.
\nonumber \\ &&
\left.
-
 %%%
 \frac{\pi}{\xi^{2 \nu}}
\left(
1- \frac{x^2}{\xi^2}
\right)^2 \,
\right.
%\nonumber \\ &&
\left.
\sum_{l=0}^{2 \nu -2}
C_{2 \nu -l-2}^{\frac{5}{2}}
\left(
\frac{x}{\xi}
\right)
%\xi
P_l
\left(
\frac{1}{\xi}
\right)
\frac{6 y^{2 \nu-l-1}}{(2 \nu-l+1)(2 \nu -l+2)}
\right\}
\,. \nonumber \\
\label{Hk_main_polarized_PS}
\ee

For even $N$ the polynomiality condition
(\ref{Polynomiality_Gluons_polarized})
require that
\be
&&
\int_0^1 dx x^{N-1} \tilde{H}^{g\,(PS)} (x,\xi,t)= \sum_{k=0 \atop  {\rm even}}^{N} \xi^k \tilde{h}^{g\,(PS)}_{N,k}(t) \nonumber \\ &&
=\xi^{N} \sum_{n=2 \atop {\rm even}}^N
  \sum_{l=0 \atop  {\rm even}}^{n}
  \tilde{B}_{nl}^{g\,(PS)}(t)
 P_l \left( \frac{1}{\xi} \right)
% \nonumber \\ && \times
 %%%%
 \frac{n\,\left( 1 + n \right) \,\left( 2 + n \right) \,\left( 3 + n \right) \,\Gamma (\frac{5}{2})\,\Gamma (N)}
  {9 \cdot 2^N\,\Gamma (1 + \frac{-n + N}{2})\,\Gamma (\frac{7}{2} + \frac{-2 + n + N}{2})}\,.
 %%%%
\label{Mellin_m_coeff_def_HPS}
\ee
The coefficients
$\tilde{h}^{g\,(PS)}_{N,k}(t)$
at powers of
$\xi$
in
(\ref{Mellin_m_coeff_def_HPS})
are given by
\be
&&
\tilde{h}^{g\,(PS)}_{N,k}(t)= %\nonumber \\ &&
\sum_{n=2 \atop \rm even}^N
\sum_{l=0 \atop \rm even}^{n}
\tilde{B}_{nl}^{g \, (PS)}(t)
%%%%%%%%%%%%%%%%%%%%%%%%%%%%%
(-1)^{\frac{k + l - N }{2}}
 %%%%%%%%%%%%%%%%%%%%%%%%%%%%%
%\nonumber \\ && \times
 \frac{\Gamma (\frac{1 - k + l + N}{2})}{ 3 \cdot 2^{k+2} \Gamma (\frac{2 + k + l - N}{2})\,\Gamma (1 - k + N)}
%%%%%%%%%%%%%%%%%%%%%%%%%%%%%%%%
\nonumber \\ && \times
\frac{n\,\left( 1 + n \right) \,\left( 2 + n \right) \,\left( 3 + n \right) \,\Gamma (N)}
  {\Gamma (\frac{2 - n + N}{2})\,\Gamma (\frac{5 + n + N}{2})}\,.
  \nonumber \\ &&
  \label{coeff_h_gluon_polarized_PS}
\ee
The expression for the pseudoscalar forward like function
$\Delta G_0^{  (PS)}$
through the pseudoscalar combination of $t$-dependent polarized gluon densities
$g^{(PS)}(y,t) \equiv \Delta g(y,t)+ \tau \Delta e^g(y,t)$
reads
\be
&&
\Delta G_0^{  (PS)}(x,t)= \frac{45}{2} x^2 \int_x^1 \frac{dy}{y^3} \Delta g^{(PS)}(y,t)
%\nonumber \\ &&
-9 x \int_x^1 \frac{dy}{y^2} \Delta g^{(PS)}(y,t)
%\nonumber \\ && -
-\frac{3}{2} \int_x^1 \frac{dy}{y} \Delta g^{(PS)}(y,t)\,.
\nonumber \\ &&
\label{G0_PS}
\ee


\begin{thebibliography}{99}
\bibitem{pioneers}
D. Mueller, D. Robaschik, B. Geyer, F.M. Dittes, and J. Horejsi,
Fortschr.~Phys. {\bf 42}, 101 (1994);\\
%%CITATION = HEP-PH 9812448;%%
%
   %%CITATION = PRLTA,78,610;%%
    A.~V.~Radyushkin,
   %``Scaling Limit of Deeply Virtual Compton Scattering,''
   Phys.\ Lett.\  B {\bf 380} (1996) 417
   [arXiv:hep-ph/9604317];\\
   %%CITATION = PHLTA,B380,417;%%
X.~D.~Ji,
   %``Deeply-virtual Compton scattering,''
   Phys.\ Rev.\  D {\bf 55} (1997) 7114
   [arXiv:hep-ph/9609381];\\
   %%CITATION = PHRVA,D55,7114;%%
    J.~C.~Collins, L.~Frankfurt and M.~Strikman,
   %``Factorization for hard exclusive electroproduction of mesons in QCD,''
   Phys.\ Rev.\  D {\bf 56} (1997) 2982
   [arXiv:hep-ph/9611433].
   %%CITATION = PHRVA,D56,2982;%%



\bibitem{pioneerJi}
X.~D.~Ji,
   %``Gauge invariant decomposition of nucleon spin,''
   Phys.\ Rev.\ Lett.\  {\bf 78} (1997) 610
   [arXiv:hep-ph/9603249].





\bibitem{GPV} K.Goeke, M.V.Polyakov and M.Vanderhaeghen, Progr. Part. Nucl.
Phys. Vol.47, No 2,  401-515 (2001) [arXiv:hep-ph/0106012].

\bibitem{Diehl}
M. Diehl, Phys. Rept. {\bf 388}, 41 (2003) [arXiv:hep-ph/0307382].

\bibitem{BelRad}
A.V. Belitsky and A.~V.~Radyushkin,
Phys. Rept. {\bf 418},  1 (2005), [arXiv:hep-ph/0504030].

%\cite{Boffi:2007yc}
\bibitem{Boffi:2007yc}
  S.~Boffi and B.~Pasquini,
  %``Generalized parton distributions and the structure of the nucleon,''
  Riv.\ Nuovo Cim.\  {\bf 30}, 387 (2007)
  [arXiv:0711.2625 [hep-ph]].
  %%CITATION = RNCIB,30,387;%%



%%%%%%%%%%%%%%%%%%%%%%%%%%%%%%%%%%%%%%%%%%%%%%%%%%%%%%%%%
% Recent status of measurements
%%%%%%%%%%%%%%%%%%%%%%%%%%%%%%%%%%%%%%%%%%%%%%%%%%%%%%%%%
%\cite{Levy:2009gy}
\bibitem{Levy:2009gy}
  A.~Levy,
  {\it  Electroproduction of Vector Mesons },
  arXiv:0907.2178 [hep-ex].
  %%CITATION = ARXIV:0907.2178;%%

 %\cite{Morrow:2008ek}
\bibitem{Morrow:2008ek}
  S.~A.~Morrow {\it et al.}  [CLAS Collaboration],
  %``Exclusive $\rho^0$ electroproduction on the proton at CLAS,''
  Eur.\ Phys.\ J.\  A {\bf 39}, 5 (2009)
  [arXiv:0807.3834 [hep-ex]].
  %%CITATION = EPHJA,A39,5;%%
%%%%%%%%%%%%%%%%%%%%%%%%%%%%%%%%%%%%%%%%%%%%%%%%%%%%%%%%%%%%

\bibitem{RDDA}
  A.~V.~Radyushkin,
  %``Nonforward parton distributions,''
  Phys.\ Rev.\  D {\bf 56}, 5524 (1997)
  [arXiv:hep-ph/9704207]; \\
%
%\bibitem{Radyushkin:1998bz}
  A.~V.~Radyushkin,
  %``Symmetries and structure of skewed and double distributions,''
  Phys.\ Lett.\  B {\bf 449}, 81 (1999)
  [arXiv:hep-ph/9810466]; \\
%
%\bibitem{RadDDandEvolution}
A.~V.~Radyushkin,
  %``Double distributions and evolution equations,''
Phys.\ Rev.\  D {\bf 59}, 014030 (1999)
[arXiv:hep-ph/9805342]; \\
%\bibitem{Musatov:1999xp}
  I.~V.~Musatov and A.~V.~Radyushkin,
  %``Evolution and models for skewed parton distributions,''
  Phys.\ Rev.\  D {\bf 61}, 074027 (2000)
  [arXiv:hep-ph/9905376].
%%%%%%%%%%%%%%%%%%%%%%%%%%%%%%%%%%%%%%%%%%%%%%%%%%%%%%%%%%%%%%%%%%%%%%%
%%%%%%%%%%%%%%%%%%%%%%%%%%%%%%%%%%%%%%%%%%%%%%%%%%%%%%%%%
% DD parametrization of the gluon GPDs
%%%%%%%%%%%%%%%%%%%%%%%%%%%%%%%%%%%%%%%%%%%%%%%%%%%%%%%%%

%\cite{Goloskokov:2005sd}
\bibitem{Goloskokov:2005sd}
  S.~V.~Goloskokov and P.~Kroll,
  %``Vector meson electroproduction at small Bjorken-x and generalized  parton
  %distributions,''
  Eur.\ Phys.\ J.\  C {\bf 42}, 281 (2005)
  [arXiv:hep-ph/0501242].
  %%CITATION = EPHJA,C42,281;%%

%\cite{Goloskokov:2007nt}
\bibitem{Goloskokov:2007nt}
  S.~V.~Goloskokov and P.~Kroll,
  %``The role of the quark and gluon GPDs in hard vector-meson
  %electroproduction,''
  Eur.\ Phys.\ J.\  C {\bf 53}, 367 (2008)
  [arXiv:0708.3569 [hep-ph]].
  %%CITATION = EPHJA,C53,367;%%

%%%%%%%%%%%%%%%%%%%%%%%%%%%%%%%%%%%%%%%%%%%%%%%%%%%%%%%%%%%%%%%%%%%%%%%%%%%%%%%%%%%%%%%%%%%%%%%%%%%%%%%%%


\bibitem{Kumericki:2007sa}
  K.~Kumericki, D.~Mueller and K.~Passek-Kumericki,
  %``Towards a fitting procedure for deeply virtual Compton scattering at
  %next-to-leading order and beyond,''
  Nucl.\ Phys.\  B {\bf 794}, 244 (2008)
  [arXiv:hep-ph/0703179].
  %%CITATION = NUPHA,B794,244;%%


%\cite{Kumericki:2008di}
\bibitem{Kumericki:2008di}
  K.~Kumericki, D.~Mueller and K.~Passek-Kumericki,
  %``Sum rules and dualities for generalized parton distributions: Is there a
  %holographic principle?,''
  Eur.\ Phys.\ J.\  C {\bf 58}, 193 (2008)
  [arXiv:0805.0152 [hep-ph]].
  %%CITATION = EPHJA,C58,193;%%

%\cite{Kumericki:2009uq}
\bibitem{Kumericki:2009uq}
  K.~Kumericki and D.~Mueller,
  {\it Deeply virtual Compton scattering at small $x_B$ and the access to the GPD
 $H$},
  arXiv:0904.0458 [hep-ph].
  %%CITATION = ARXIV:0904.0458;%%
%%%%%%%%%%%%%%%%%%%%%%%%%%%%%%%%%%%%%%%%%%%%%%%%%%%%%%%%%%%%%%%%%%%%%%%%%%%%%%%%%%%%


%
\bibitem{Polyakov:2002wz}
  M.~V.~Polyakov and A.~G.~Shuvaev,
  {\it  On 'dual' parametrization of generalized parton distributions },
  arXiv: hep-ph/0207153.

\bibitem{Tomography}
  M.~V.~Polyakov,
  %``Tomography for amplitudes of hard exclusive processes,''
  Phys.\ Lett.\  B {\bf 659}, 542 (2008)
  [arXiv:0707.2509 [hep-ph]].


\bibitem{Ji:1998pc}
  X.~D.~Ji,
 % ``Off-forward parton distributions,''
  J.\ Phys.\ G {\bf 24}, 1181 (1998)
  [arXiv:hep-ph/9807358].
  %%CITATION = JPHGB,G24,1181;%%

%Problems with GPDs
%\cite{Guidal:2007cw}
\bibitem{Guidal:2007cw}
  M.~Guidal and S.~Morrow,
 {\it  Exclusive $\rho^0$ electroproduction on the proton : GPDs or not GPDs ?},
  arXiv:0711.3743 [hep-ph].
  %%CITATION = ARXIV:0711.3743;%%

\bibitem{Polyakov:2007rw}
  M.~V.~Polyakov,
 {\it   Educing GPDs from amplitudes of hard exclusive processes},
  arXiv:0711.1820 [hep-ph].




%\cite{Moiseeva:2008qd}
\bibitem{Moiseeva:2008qd}
  A.~M.~Moiseeva and M.~V.~Polyakov,
  {\it  Dual parameterization and Abel transform tomography for twist-3
  DVCS,}
  arXiv:0803.1777 [hep-ph].

%\cite{Polyakov:2008xm}
%\bibitem{Polyakov:2008xm}
 % M.~V.~Polyakov and M.~Vanderhaeghen,
 % {\em ``Taming Deeply Virtual Compton Scattering,''}
 % arXiv:0803.1271 [hep-ph].

%\cite{SemenovTianShansky:2008mp}
\bibitem{ForwardLikeF_KS}
 K.~M.~Semenov-Tian-Shansky,
  %``Forward-like functions for dual parametrization of GPDs from nonlocal
  %chiral quark model,''
  Eur.\ Phys.\ J.\  A {\bf 36}, 303 (2008)
  [arXiv:0803.2218 [hep-ph]].
  %%CITATION = EPHJA,A36,303;%%

\bibitem{DualVSRad}
 M.~V.~Polyakov and K.~M.~Semenov-Tian-Shansky,
  %``Dual parametrization of GPDs versus double distribution Ansatz,''
  Eur.\ Phys.\ J.\  A {\bf 40}, 181 (2009)
  [arXiv:0811.2901 [hep-ph]].

\bibitem{Polyakov:1998ze}
  M.~V.~Polyakov,
  %``Study of two-pion light-cone distribution amplitudes in the resonance
  %region and at low energies,''
  Nucl.\ Phys.\  B {\bf 555}, 231 (1999)
  [arXiv:hep-ph/9809483].



  \bibitem{Dolen:1967zz}
  R.~Dolen, D.~Horn and C.~Schmid,
  %``Prediction of Regge Parameters of rho Poles from Low-Energy pi N Data,''
  Phys.\ Rev.\ Lett.\  {\bf 19}, 402 (1967).

   \bibitem{Alfaro_red_book}
 V.~de Alfaro, S.~Fubini, G.~Furlan, C.~Rossetti,
 {\it Currents in Hadron Physics}
 (North-Holland, Amsterdam, 1973).


%\cite{SemenovTianShansky:2007hv}
\bibitem{SemenovTianShansky:2007hv}
  K.~M.~Semenov-Tian-Shansky, A.~V.~Vereshagin and V.~V.~Vereshagin,
  %``Bootstrap and the physical values of $\pi N$ resonance parameters,''
  Phys.\ Rev.\  D {\bf 77}, 025028 (2008)
  [arXiv:0706.3672 [hep-ph]].
  %%CITATION = PHRVA,D77,025028;%%

%\cite{Braun:2003rp}
\bibitem{Braun:2003rp}
  V.~M.~Braun, G.~P.~Korchemsky and D.~Mueller,
  %``The uses of conformal symmetry in QCD,''
  Prog.\ Part.\ Nucl.\ Phys.\  {\bf 51}, 311 (2003)
  [arXiv:hep-ph/0306057].
  %%CITATION = PPNPD,51,311;%%


\bibitem{Belitsky:1997pc}
  A.~V.~Belitsky, B.~Geyer, D.~Mueller and A.~Schafer,
  %``On the leading logarithmic evolution of the off-forward distributions,''
  Phys.\ Lett.\  B {\bf 421}, 312 (1998)
  [arXiv:hep-ph/9710427].



\bibitem{Shuvaev:1999fm}
  A.~Shuvaev,
  %``Solution of the off-forward leading logarithmic evolution equation  based
  %on the Gegenbauer moments inversion,''
  Phys.\ Rev.\  D {\bf 60}, 116005 (1999)
  [arXiv:hep-ph/9902318].

\bibitem{Kivel:1999wa}
  N.~Kivel and L.~Mankiewicz,
  %``Conformal string operators and evolution of skewed parton  distributions,''
  Nucl.\ Phys.\  B {\bf 557}, 271 (1999)
  [arXiv:hep-ph/9903531].

\bibitem{Manashov:2005xp}
  A.~Manashov, M.~Kirch and A.~Schafer,
  %``Solving the leading order evolution equation for GPDs,''
  Phys.\ Rev.\ Lett.\  {\bf 95}, 012002 (2005)
  [arXiv:hep-ph/0503109].


\bibitem{Mueller:2005ed}
  D.~Mueller and A.~Schafer,
  %``Complex conformal spin partial wave expansion of generalized parton
  %distributions and distribution amplitudes,''
  Nucl.\ Phys.\  B {\bf 739}, 1 (2006)
  [arXiv:hep-ph/0509204].
  %%CITATION = NUPHA,B739,1;%%

%\cite{Kumericki:2007sa}







%\cite{Diehl:2007jb}
\bibitem{Diehl:2007jb}
  M.~Diehl and D.~Y.~Ivanov,
  %``Dispersion representations for hard exclusive processes,''
  Eur.\ Phys.\ J.\  C {\bf 52}, 919 (2007)
  [arXiv:0707.0351 [hep-ph]].
  %%CITATION = EPHJA,C52,919;%%


\bibitem{Munczec:63}
 H.~Munczec, Nuovo Cimento, {\bf{29}}, 1175 (1963).

\bibitem{Alfaro:67}
 V De Alfaro, S Fubini, C Rossetti, G Furlan,
 % Superconvergence and current algebra,
 Annals of Physics Volume 44, Issue 2 , 165 (1967).


\bibitem{Martin:2002dr}
  A.~D.~Martin, R.~G.~Roberts, W.~J.~Stirling and R.~S.~Thorne,
  %``NNLO global parton analysis,''
  Phys.\ Lett.\  B {\bf 531}, 216 (2002)
  [arXiv:hep-ph/0201127].



%%%%%%%%%%%%%%%%%%%%%%%%%%%%%%%%%%%%%%%%%%%%%%%%
\bibitem{Dispersion}
 O.~V.~Teryaev,
{\it Analytic properties of hard exclusive amplitudes},
  arXiv:hep-ph/0510031; \\
  I.~V.~Anikin and O.~V.~Teryaev,
  %``Dispersion relations and subtractions in hard exclusive processes,''
  Phys.\ Rev.\  D {\bf 76}, 056007 (2007)
  [arXiv:0704.2185 [hep-ph]]; \\
 % M.~Diehl and D.~Y.~Ivanov,
  %``Dispersion representations for hard exclusive processes,''
  %Eur.\ Phys.\ J.\  C {\bf 52}, 919 (2007)
  %[arXiv:0707.0351 [hep-ph]]; \\
M.~Diehl and D.~Y.~Ivanov,
  {\it Dispersion representations for hard exclusive reactions},
  arXiv:0712.3533 [hep-ph].




%\cite{Martin:2008tm}
\bibitem{Martin:2008tm}
  A.~D.~Martin, C.~Nockles, M.~G.~Ryskin, A.~G.~Shuvaev and T.~Teubner,
%{\em  `` Generalized parton distributions at small $x$,''}
Eur.\ Phys.\ J.\  C {\bf 63}, 57 (2009)
 [arXiv:0812.3558 [hep-ph]].
  %%CITATION = ARXIV:0812.3558;%%



%\cite{Shuvaev:1999fm}
%\bibitem{Shuvaev:1999fm}
 % A.~Shuvaev,
  %``Solution of the off-forward leading logarithmic evolution equation  based
  %on the Gegenbauer moments inversion,''
  %Phys.\ Rev.\  D {\bf 60}, 116005 (1999)
  %[arXiv:hep-ph/9902318].
  %%CITATION = PHRVA,D60,116005;%%

%\cite{Martin:2008tm}

\bibitem{Guidal:2004nd}
  M.~Guidal, M.~V.~Polyakov, A.~V.~Radyushkin and M.~Vanderhaeghen,
  %``Nucleon form factors from generalized parton distributions,''
  Phys.\ Rev.\  D {\bf 72}, 054013 (2005)
  [arXiv:hep-ph/0410251].


\bibitem{Gelfand}
I.~M.~Gelfand and G.~E.~Shilov, {\it Generalized Functions}, Vol. I
(Academic Press, New York, 1964).





%\cite{:2007cz}
\bibitem{:2007cz}
  F.~D.~Aaron {\it et al.}  [H1 Collaboration],
  %``Measurement of Deeply Virtual Compton Scattering and its t-dependence at
  %HERA,''
  Phys.\ Lett.\  B {\bf 659}, 796 (2008)
  [arXiv:0709.4114 [hep-ex]].
  %%CITATION = PHLTA,B659,796;%%

%\cite{Freund:2002qf}
\bibitem{Freund:2002qf}
  A.~Freund, M.~McDermott and M.~Strikman,
  %``Modelling generalized parton distributions to describe deeply virtual
  %Compton scattering data,''
  Phys.\ Rev.\  D {\bf 67}, 036001 (2003)
  [arXiv:hep-ph/0208160].
  %%CITATION = PHRVA,D67,036001;%%

  %\cite{Kivel:2000fg}

\bibitem{Kivel:2000fg}
  N.~Kivel, M.~V.~Polyakov and M.~Vanderhaeghen,
  %``DVCS on the nucleon: Study of the twist-3 effects,''
  Phys.\ Rev.\  D {\bf 63}, 114014 (2001)
  [arXiv:hep-ph/0012136].
  %%CITATION = PHRVA,D63,114014;%%




%\cite{Polyakov:2002yz}
\bibitem{Polyakov:2002yz}
  M.~V.~Polyakov,
  %``Generalized parton distributions and strong forces inside nucleons and
  %nuclei,''
  Phys.\ Lett.\  B {\bf 555}, 57 (2003)
  [arXiv:hep-ph/0210165].
  %%CITATION = PHLTA,B555,57;%%

%\cite{Kumericki:2009ji}
\bibitem{Kumericki:2009ji}
  K.~Kumericki and D.~Mueller,
  {\it DVCS and the skewness effect at small x},
  arXiv:0907.1207 [hep-ph].
  %%CITATION = ARXIV:0907.1207;%%


\bibitem{Ryzhik}
I.~S.~Gradshteyn, I.~M.~Ryzhik,
{\it  Table of Integrals, Series, and Products }
(Academic Press, 2000).

\end{thebibliography}
\end{document}